\documentclass{article}

\usepackage{arxiv}

\usepackage[utf8]{inputenc} 
\usepackage[T1]{fontenc}    
\usepackage{hyperref}       
\usepackage{url}            
\usepackage{booktabs}       
\usepackage{amsfonts}       
\usepackage{nicefrac}       
\usepackage{microtype}      
\usepackage{lipsum}		
\usepackage{graphicx}
\usepackage{tabularx}
\usepackage{caption} 
\usepackage[labelfont=bf]{caption}
\usepackage[
singlelinecheck=false 
]{caption}
\usepackage[numbers]{natbib}
\usepackage{pdfpages}

\title{Reimagining Speech: A Scoping Review of Deep Learning-Powered Voice Conversion}

\author{{Anders R. Bargum} \\
        Khora \& Heka VR\\
	Aalborg University\\
	Copenhagen, Denmark \\
	\texttt{arba@create.aau.dk} \\
    \And
	{Stefania Serafin} \\
	Multisensory Experience Lab\\
	Aalborg University\\
	Copenhagen, Denmark \\
	\texttt{sts@create.aau.dk} \\
    \And
	{Cumhur Erkut} \\
	Multisensory Experience Lab\\
	Aalborg University\\
	Copenhagen, Denmark \\
	\texttt{cer@create.aau.dk} \\
}

\date{}

\renewcommand{\headeright}{}
\renewcommand{\undertitle}{}
\renewcommand{\shorttitle}{}

\hypersetup{
pdftitle={A template for the arxiv style},
pdfsubject={q-bio.NC, q-bio.QM},
pdfauthor={David S.~Hippocampus, Elias D.~Striatum},
pdfkeywords={First keyword, Second keyword, More},
}

\begin{document}
\maketitle

\begin{abstract}
    Research on deep learning-powered voice conversion (VC) in speech-to-speech scenarios is getting increasingly popular. Although many of the works in the field of voice conversion share a common global pipeline, there is a considerable diversity in the underlying structures, methods, and neural sub-blocks used across research efforts. Thus, obtaining a comprehensive understanding of the reasons behind the choice of the different methods in the voice conversion pipeline can be challenging, and the actual hurdles in the proposed solutions are often unclear. To shed light on these aspects, this paper presents a scoping review that explores the use of deep learning in speech analysis, synthesis, and disentangled speech representation learning within modern voice conversion systems. We screened 621 publications from more than 38 different venues between the years 2017 and 2023, followed by an in-depth review of a final database consisting of 123 eligible studies. Based on the review, we summarise the most frequently used approaches to voice conversion based on deep learning and highlight common pitfalls within the community. Lastly, we condense the knowledge gathered, identify main challenges and provide recommendations for future research directions.
\end{abstract}


\section{Introduction}
Voice transformations (VT) describe the act of controlling non-linguistic characteristics of speech such as the quality or the individuality of a vocal signal \cite{Stylianou}. The expression "transformation" is used as an umbrella term referring to the modifications made in a speech-to-speech scenario where an application or a technical system is used to map, modify or modulate specific characteristics of a voice, be it its pitch, timbre or prosody.

A sub-task of voice transformation is the topic of voice conversion (VC). More specifically, VC seeks to render an utterance from one speaker to sound like that of a target speaker and it has in the past decade become a prominent research subject within the field of artificial intelligence (AI). Most commonly, voice conversion refers to the process of changing the properties of speech, such as voice identity, emotion, language or accent and the process has in the past years made a major impact on several real-life applications such as personalized speech
synthesis, communication aids for speech impaired or simple voice mimicry. It should be noted that VC also is used to describe the conversion procedure of a text-to-speech (TTS) pipeline in which a user chooses specific speaker characteristics that the utterances should sound like \cite{overviewSisman}. This review considers only the former definition. 

The voice conversion pipeline can be divided into three main stages: 1) \textit{the speech analysis} stage where linguistic content and speaker timbral information are extracted individually and disentangled from each other, 2) \textit{the mapping stage} where the decomposed information is transferred towards an intermediate representation that matches the qualities of a specific target speaker, this is often done using a decoder or a generator. And, 3) \textit{the reconstruction and synthesis stage} where the decoded, intermediate representation is processed and re-synthesised into the time-domain using a vocoder \cite{overviewWalczyna}. Assuming that parallel data i.e. the same utterances spoken by both the input and target speaker, is available, the above conversion stages have traditionally been carried out through the procedure of "spectrum mapping". This has most commonly been accomplished using statistical models and signal processing techniques such as Gaussian mixture models (GMM) \cite{Stylianou2} and non-negative matrix factorization (NMF) \cite{Wu2013ExemplarbasedVC} whereas synthesis has been achieved through techniques based on the inverse fourier transform and pitch synchronous overlap-add (PSOLA) methods \cite{psola}. However, the advancements in deep learning have facilitated the development of non-parallel and end-to-end training methods, while promoting both new mapping functions and vocoding techniques. Consequently, the advancements have led to substantial improvements in the quality and fidelity of the conversion outcomes and deep learning has thus become 'a new standard' for carrying out voice conversion today \cite{overviewSisman}.

Currently, few works have reviewed the field of voice conversion techniques. In \cite{overviewSisman}, the authors give a comprehensive overview of the history of voice conversion technology and identify common deep learning methods used in the literature. Their work is extended by \cite{overviewWalczyna}, who focuses on the description of frequently used methods and models - both in terms of their analysis and mapping stages and their synthesis paradigms. Our work accompanies the above-mentioned reviews, and should be seen as complementary. However, rather than giving an extensive summary of the field and its history, we direct the analysis into an investigation of the main problem areas addressed in current VC research. We do this to further highlight popular topics in the area, discuss challenges encountered and investigate future directions of the research.

More specifically, we present a review of \underline{123} papers published between 2017 and 2023. We do this to firstly understand which disentanglement methods, training techniques and neural vocoders are most frequently used and secondly, to emphasise the areas in which VC needs further development. Since voice conversion approaches generally differ by their training configurations, we additionally provide a quantitative insight into the employed training specifics including the incorporated loss functions and feature extraction algorithms utilized. The papers included in this review have been selected using a set of inductively identified keywords and they have all undergone a filtering process as well as a three-stage screening procedure. We review the work and extract data based on a codebook of 14 codes addressing topics such as \textit{research direction}, \textit{training specifications} and \textit{datasets used} as well as \textit{encoders, decoders and vocoders} utilised. The extracted data forms the basis for both a thematic analysis and statistical synthesis that we use to summarise the current state of the art giving a comprehensive overview of the field. 

In summary, this scoping review makes the following contributions: Firstly, we provide an intuitive overview of the voice conversion pipeline in order for the reader to understand the topics addressed in the forthcoming work. Secondly, we furnish a dataset comprising reviewed papers, employed codes, numerical representations and keywords, to elucidate the instruments, techniques, and constituent components used in the analysis of the chosen studies. Thirdly, we present a graphical representation depicting the distributions of work. In conclusion, we undertake a thorough analysis and provide a comprehensive summary of the discrete elements utilized within a traditional voice conversion pipeline. Our primary emphasis is here directed towards specific areas of development notably; style transfer, identity conversion, and the exploration of potential interpretability.

\section{Background} \label{sec:background}
Before presenting the methodology, we introduce general terminology and describe the sub-blocks and typical flow of a traditional deep learning-based voice conversion pipeline. The following Section serves as a brief overview that covers important aspects needed for the reader to understand the codes and descriptions employed throughout the coding, analysis and synthesis procedures.

\subsection{The Voice Conversion Pipeline}
\begin{figure}[!htbp]  
\centering
\includegraphics[width=1.0\textwidth]{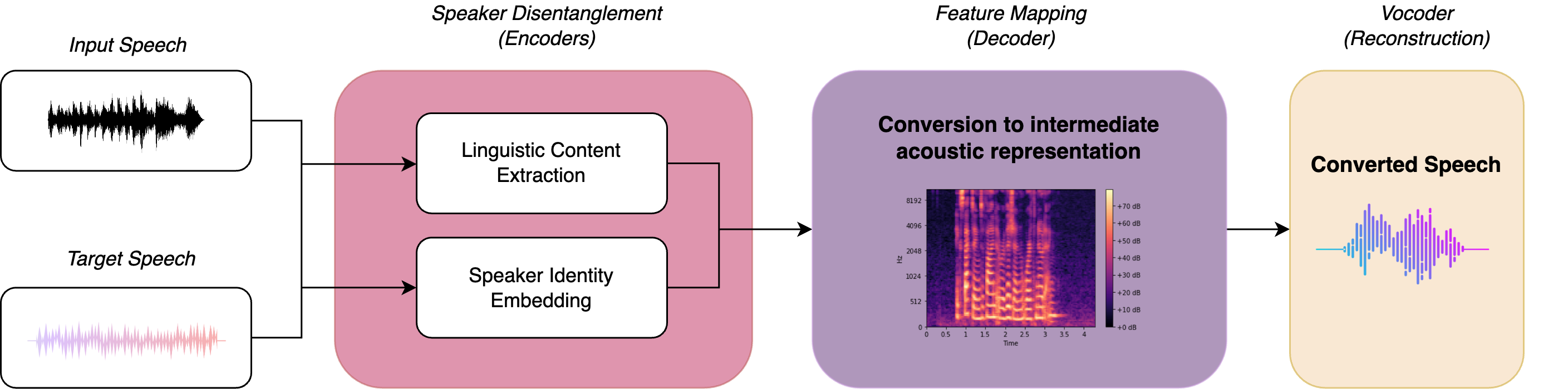}
\caption{\textit{Illustration of a traditional deep learning based voice conversion pipeline and its three main stages.
}}\label{fig:pipeline}
\end{figure}

Research on speech analysis and synthesis has been conducted since 1922 where Stewart \cite{Stewart1922AnEA} commented that the difficult problems involved in artificial voice production consist in the manipulation of the apparatus producing it rather than the actual production of the speech itself. In order to understand and implement both speech analysis, production and manipulation blocks in such an apparatus we can utilise the ability to characterize speech by different factors. Firstly we can define speech by its \textit{linguistic factors}, which are reflected in sentence structure, lexical corpus and idiolect e.g. words and speech habits. Secondly, we can divide it into \textit{supra-segmental factors}, which are the prosodic attributes of speech such as intonation, stress and rhythm, and \textit{segmental factors}, which are related to speaker identity and timbre such as the spectrum, spectral envelope and formants \cite{overviewSisman}. These aspects can be interchanged and mapped in various ways, however, most deep learning-powered VC approaches train a conversion model that transforms the segmental factors extracted from an utterance of a source speaker in order to match that of a target speaker, all while the linguistics remain unchanged. We provide an illustration of the typical voice conversion pipeline including analysis, mapping, and reconstruction modules in Figure \ref{fig:pipeline} and divide the stages depicted into even more specific tasks.
\newpage

Most VC pipelines start by extracting information about linguistic content and speaker identity individually. In the field of VC we denote the timbral and segmental aspects of speaker identity as \textit{speaker-dependent} features i.e. features that capture the specific vocal characteristics of an individual speaker, while the remaining linguistics are denoted as \textit{speaker-independent} features i.e. features that describe spoken content universal to any spoken language. Once the speaker-dependent and speaker-independent factors are extracted, the voice conversion process can be recast as a style transfer problem, where speech characteristics are regarded as styles, and speaker-independent factors are regarded as domains \cite{qian2019autovc}. The key idea behind the style transfer formulation is to achieve full disentanglement between the styles and domains from which one can manipulate and/or replace the embeddings. Here, one can incorporate timbral information from the input itself, which will result in pure reconstruction, or involve the speaker embedding from other speakers, which will result in the actual conversion, or style transfer, matching the characteristics of the speaker inserted \cite{overviewWalczyna}. In contrast to the parallel conversion process, wherein the transformation primarily involves spectrum mapping, style transfer in a non-parallel setting presents one key challenge: the absence of alignment in the non-parallel data. One approach to addressing this challenge involves phonetic posteriograms (PPGs) that establish an intermediary phonetic representation of both the source and target \cite{PPG}. Nevertheless, as will be subsequently discussed, several alternative methods are available.

In the literature, numerous models have been proposed to both extract and map abovementioned features. \textit{Generative adversarial networks} (GANs) and \textit{(variational) auto-encoders} (AEs/VAEs) are particularly popular choices. In \cite{GAN-VC} and \cite{GAN-SVC} they as an example utilise traditional GAN-based training schemes with timbre representation losses used to train a system that can extract and match timbre characteristics of many speakers. Differently, in \cite{dhar2021adaptive} the generator of a traditional GAN network is extended with adaptive learning using a dense residual network (DRN) to enhance the feature learning ability i.e. speaker generalization, of their proposed model. In \cite{cyclegan}, adversarial weight-training paradigms are used to map the feature to more realistic representations by creating balance in the, sometimes, unstable nature of a GAN. This is done by giving more attention to samples that fool the discriminator allowing the generator to learn more from “true” samples rather than “fake” ones. The study further imparts an inductive bias by utilizing spectral envelopes as input data for the generator. By doing this they limit the conversion task to subtle adjustments of the spectral formants, promoting ease of learning in the often challenging training scheme of GANs.

Contrary to the GAN-based approaches, the authors in \cite{du2022disentanglement, DEEPA, tang2022avqvc} utilise the benefits of representation learning in AEs and VAEs. By doing so their models are forced to separate linguistic and timbre information through an information bottleneck. Nonetheless, the size and features used to encode the latent spaces as well as the explicit control offered by these models differ. In \cite{du2022disentanglement}, traditional content and speaker embeddings are used to condition the decoder, which in turn produces a mel-spectrogram to be used by the vocoder. Contrary, the work by \cite{DEEPA} focuses on learning a latent representation from which the decoder can create harmonic and noise components matching that of the target speech. Lastly, \cite{tang2022avqvc} encodes a broader range of information, including speaker, content, style, and pitch (F0), making it easier to force disentanglement and interchange chosen features in the conversion process. Despite the improved efficiency of AE and VAE-based representations, Wu et al. \cite{wu2020vqvc} note that it in some cases may produce imperfect disentanglement harming the quality of the output speech. This happens as weaknesses in any intermediate and individual module may cascade errors in the overall system. To address this, \cite{wu2020vqvc} further extend the auto-encoder-based VC framework with a U-Net architecture and force a strong information bottleneck using vector quantization (VQ) on the latent vectors \cite{wu2020vqvc}. The latter is done to prevent the U-Net from overfitting on the reconstruction task and will later be shown to be a popular choice in regularizing the latent space (see Section \ref{sec:disentangling}).

The recent rise in language models (LMs) has additionally shown promising results for voice conversion and feature mapping. In \cite{wang2023lmvc}, they propose 'LM-VC' in which the usual embeddings are substituted by tokens known from language representations. Here a two-stage masked language model generates coarse acoustic tokens for recovering both the source linguistic content and the target speaker’s timbre. The approach is shown to outperform competitive systems, both for speech naturalness and speaker similarity, however, the model is restricted to the use of well-known tokenizers, which often contain millions of parameters \cite{hsu2021hubert}. Voice conversion systems relying on language models (LM) are therefore inherently intricate, lacking interpretability, and demonstrate inefficiency during inference. They may therefore not necessarily always contribute positively to the voice conversion process. This will further be discussed in Section \ref{sec:structures}.

As noted in Figure \ref{fig:pipeline}, the speech is finally reconstructed by synthesising the intermediate acoustic representation back into the time-domain. While this classically has been achieved by the Griffin-Lim algortihm or inverse fourier transforms, current work utilises neural vocoders such as the WaveNet \cite{oord2016wavenet} or the HiFi-GAN \cite{lian_towards_2022}. These processes are known for their high fidelity and robustness toward modifications in the intermediate representations - aspects that are crucial for high output quality. The use and inclusion of vocoders will be examined in Section \ref{sec:vocoders}

Besides complexity, freedom and modularity, the introduction of deep learning signifies a departure from the conventional analysis-mapping-reconstruction pipeline. Above mentioned techniques may therefore all be trained in an end-to-end manner, substituting each sub-task with other neural processes either from similar VC work or from completely different speech processing fields. As we advance, the subsequent sections will navigate deeper into the intricacies of these techniques. The forthcoming sections will serve as a bridge to the results, offering a granular perspective on the approach taken when choosing and extracting our data.

\section{Method}
The decision to undertake a scoping review in the domain of deep learning-powered voice conversion has been informed by the transformative nature of deep learning, as emphasized in \cite{overviewSisman}, where differentiable techniques have shifted the paradigm away from the traditional analysis-mapping-reconstruction pipeline. This shift enables end-to-end training, providing both flexibility and improved target matching, however, challenges arise depending on the methods incorporated. We therefore seek to survey current techniques used and anticipate future challenges. Unlike other review types, scoping reviews aim to "identify and map the available evidence" \cite{scopingReview1} and thus focus on the quality and quantity of key features, rather than answering specific questions \cite{scopingReview2}. While no formal quality assessment is needed in a scoping review, Colquhoun et al. \cite{COLQUHOUN20141291} recommend following a few simple guidelines to ensure consistency in the analysis and synthesis phases. As suggested by Colquhoun et al., we have chosen to follow Arksey and O’Malley's framework stages for the conduct of scoping reviews combined with the Levac et al. enhancements \cite{COLQUHOUN20141291}. In the review process we thus take the following steps: 1) Identify the research question, 2) Identify relevant studies, 3) Select and screen relevant studies, 4) Chart the gathered data, 5) Collate, summarise and report the results \cite{COLQUHOUN20141291}. We furthermore integrate the PRISMA for scoping review (PRISMA ScR) checklist \cite{prisma} into the guidelines, ensuring consistency and objectivity throughout the iterative reviewing process. The latter aspects have been specifically important as the reviewing process has been carried out by fewer authors than recommended, making the synthesis and results receptive to subjective bias.

\subsection{Research Questions}
\label{sec:rq}
We guide our review of deep learning-based voice conversion by the following research objectives; \textit{1) identify the current state of the art in the field of deep learning-based voice conversion}, \textit{2) identify the commonly used tools, techniques, and evaluation methods in deep learning-based voice conversion research}, \textit{3) gain a comprehensive understanding of the requirements and existing gaps in different voice conversion frameworks}. To accomplish these objectives, our review will address the following research questions: 
\begin{itemize}
    \item What are the fundamental components that comprise high-fidelity voice conversion pipelines?
    \item What are the primary areas of concern addressed in research on voice conversion and which challenges are they trying to solve?
\end{itemize}
More specifically, our review will encompass an examination of research findings and standardized methodologies in the domain of voice conversion. Furthermore, we aim to provide a quantitative analysis of the approaches employed at each stage of the conversion pipeline, clarifying the rationale behind the selection and application of these techniques.

\subsection{Keyword identification}
\begin{table*}[t]
\caption{\textit{Keyword list for literature search. The search is done by placing an AND between the first and last column and an OR between the rows of each column}}\label{tab:keyword-list}
\footnotesize
\begin{tabularx}{\textwidth}{X | X | X | X}
{\textbf{Keyword 1}} & {\textbf{Keyword 2 (Method)}} & {\textbf{Keyword 3 (Subtopic)}} & {\textbf{Keyword 4 (Feature)}}\\
\hline
& Deep learning & Style transfer &\\
& Convolutional NNs & Speech Synthesis & Pitch\\
& Generative adversarial & Disentanglement & Timbre\\
Voice Conversion & Unsupervised & Vocoder & Formant \\
& Adversarial (training*) & Zero-shot & Energy \\
& Self-supervised & Conditioning &  Dynamics \\
& Vector quantization & End-to-end & Prosody \\
& Autoencoder & Speaker embedding & \\
\hline
\end{tabularx}
\end{table*}
Relevant studies were retrieved using research-specific keywords. We identified the keywords using a data-driven approach where one main keyword guided the search for related keywords. We did this to ensure objectivity and overcome limitations regarding knowledge gaps or biases towards terms we would use to describe the research objectives at hand. To find deep learning terms connected to voice conversion, we searched for relevant keywords in the 2022 proceedings of two machine learning and audio-related conferences (ICASSP\footnote{International Conference on Acoustics, Speech, and Signal Processing: \url{https://ieeexplore.ieee.org/xpl/conhome/9745891/proceeding}, accessed 05.07.2023} and NeurIPS\footnote{Advances in Neural Information Processing Systems:\\ \url{https://papers.nips.cc/paper\_files/paper/2022}, accessed 05.07.2023}) using the main keyword 'voice conversion'. For all papers retrieved, author A screened the relevance of the results by reading the full title and abstracts where-after the global keyword list was updated by the author keywords from each paper. In total, 18 relevant papers in the realm of voice conversion and deep learning were found. In these papers, 16 unique keywords were repeated more than once. The complete list is shown in Table 1 with each keyword sorted into sub-topics. We specifically excluded the keyword "text" to avoid searching for studies with a focus on TTS-based voice conversion. Lastly, we added feature-based keywords such as pitch, timbre, formant, energy and dynamics to further force audio domain-specificity. 

\subsection{General Search}
We ensured that the voice conversion content was limited to deep learning techniques using AND operators between the first column and the last column of the keyword list, while OR operators were used between the remaining columns and rows. This meant that articles searched for, all contained the keyword 'voice conversion' in addition to popular deep learning methods, subtopics and features. We queried the Scopus\textregistered\footnote{Scopus: \url{https://www.scopus.com/search}, accessed 23.10.2023} and the Web Of Science\footnote{Web of Science: \url{https://www.webofscience.com}, accessed 23.10.2023} databases due to the former's high scientific journal rankings and the latter's indexing of conferences such as ICASSP and NeurIPS. We searched for the keywords in all material of the Scopus archive and limited the search to title, abstract and keywords for the Web of Science archive. Initially, 621 papers were retrieved, 422 coming from Scopus and 199 coming from Web of Science. An additional filtering process was carried out in which the search was limited to 6 years from 2017-2023, and filtered to leave out any reviews, surveys, book chapters, letters and thesis papers. The filtering resulted in 573 final papers. Deep learning is a quickly growing field and the above-mentioned process was chosen to display its impact on the current state of the art within voice conversion.

We adopted a three-stage screening phase for the remaining literature; 1) \textit{the title-phase}, 2) \textit{the abstract-phase} and 3) the \textit{full text-phase}. The first phase started with removing duplicated publications. Thereafter author A inspected the remaining papers based on their title alone. Two main criteria were used in this phase: Firstly, the works should be written in English and secondly, the title should relate to the aspect of voice conversion, e.g. material on regulation, detection or anti-spoofing was removed. This stage served as an additional filtering phase, taking account of any mistakes in the first filtering stage which as an example could have been caused by inaccurate metadata. Phase 1 discarded 357 papers mostly due to the lack of relevance. In the second phase, abstracts of the remaining 216 works were read and evaluated according to several exclusion/eligibility criteria (EC) which were iteratively created throughout phase 1: 

\begin{itemize}
    \item[\textbf{EC1}] \textit{Modality:} The main focus of the paper/article is on other modalities e.g. video information or text-to-speech systems. Only direct voice conversion using speaker-to-speaker or reconstruction methods (audio-to-audio) should be included.
    \item[\textbf{EC2}] \textit{Purpose:} The paper/article has a bigger purpose than traditional voice conversion. E.g. it aims to achieve speaker recognition, and identification, recreate pathological voices or convert whispers and screams.
    \item[\textbf{EC3}] \textit{Synthesis:}
    The paper deals with speech synthesis/neural vocoding only.
    \item[\textbf{EC4}] \textit{Method:} The paper does not include any deep learning techniques (GANs, AEs, VAEs, RNNs, Attention Mechanisms etc.).
    \item[\textbf{EC5}] \textit{Singing Voice:} The system is focused on singing voice synthesis or conversion.
    \item[\textbf{EC6}] \textit{Lack of VC information:} The paper lacks general information on the voice conversion process e.g. it focuses on evaluation methods. 
\end{itemize}

\begin{figure*}[t]  
\centering
\includegraphics[width=1.0\textwidth]{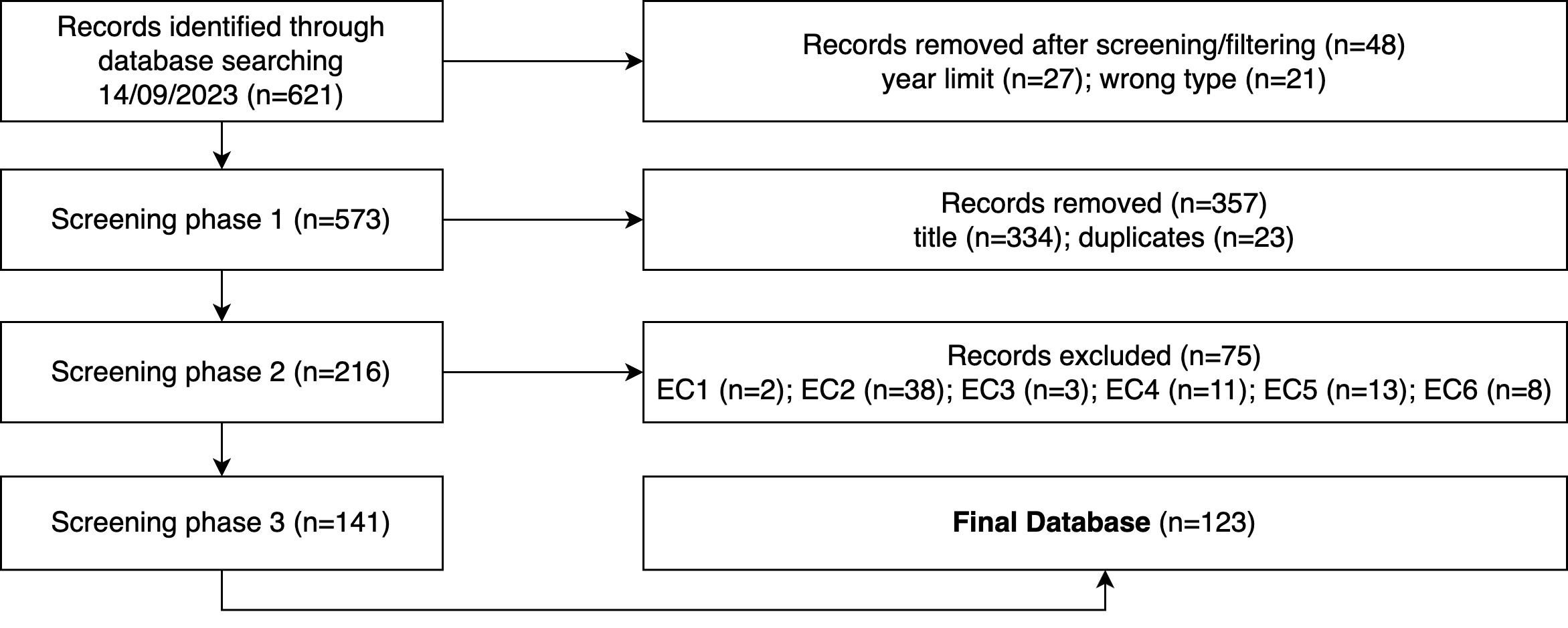}
\caption{\textit{Prisma-ScR chart documenting the retrieval process of identified sources of evidence used for data extraction and analysis.}
}\label{fig:prisma}
\end{figure*}

During phase 2, 75 papers were excluded leaving behind 141 papers used for full text analysis and coding. During the code extraction phase, we found another 18 papers to be either irrelevant or non-retrievable. This means a final pool of 123 papers has been included in this review. We picture the full source selection process using the PRISMA diagram for scoping reviews (PRISMA-ScR) \cite{prisma} in Figure \ref{fig:prisma}.

\subsection{Data Items and Code Book}
The 123 papers were carefully read, analysed and summarised. We charted the papers based on three main topics: 1) \textit{Research objective and contributions}; what was the goal of the authors and what did they achieve? 2) \textit{Methods and techniques used}; How did the authors achieve their goal and which deep learning methods and intermediate features were used? 3) \textit{Evaluation and miscellaneous}; How did the authors evaluate their work, what were the results and did they apply any manipulation techniques (eg. augmentation, perturbation, regularization)? Data items related to each coding topic can be seen in Table \ref{tab:code-list} while the complete code book can be found in Appendix A. 

\begin{table*}[h]

\caption{\textit{Codes used for data extraction}}\label{tab:code-list}
\footnotesize
\begin{tabularx}{\textwidth}{X | X | X | X}
{\textbf{Research Objective}} & {\textbf{Deep Learning Methods}} & {\textbf{Evaluation \& Misc.}} & {\textbf{Training Specs}}\\
\hline
\textit{C1} Category & \textit{C4} Global Structure & \textit{C8} Objective Evaluation & \textit{C12} Loss Function \\
\textit{C2} Main Goal & \textit{C5} Analysis Features & \textit{C9} Subjective Evaluation & \textit{C13} Optimizer Used \\
\textit{C3} Contributions & \textit{C6} Encoders used & \textit{C10} Dataset Used & \textit{C14} Sampling Rate \\ & \textit{C7} Vocoder Type & \textit{C11} Manipulation & \\
\hline
\end{tabularx}
\end{table*}

\subsection{Delimitation}
While scoping reviews by default include all available evidence regardless of methodological quality \cite{arksey}, it was chosen to limit the search to full journal articles and paper proceedings only. This was decided to guarantee a manageable evidence outcome. Additionally, it is worth emphasizing the number of authors involved in the review process, as this may have had an impact on certain aspects of the review. For instance, the assessments have only undergone external review by one individual. Nevertheless, we endeavoured to mitigate potential selection bias through an objective and meticulously chosen keyword-selection process and to address interpretation bias through ongoing, rigorous coding discussions and meetings.

\section{Results}
The following Section presents a combined analysis and discussion of the results of this review. Firstly, we present a summary of the papers’ research directions and distributions of deep learning methods used. Secondly, we provide an exposition of the papers' relationship with the traditional voice conversion pipeline explained earlier, substantiating our analysis with quantitative data and illustrative materials. Thirdly, the main problem areas addressed in voice conversion research are outlined counting a discussion of topics such as; interpretability, prevalent conditioning features, and the encountered challenges in integrating explicit control mechanisms. As we aim to reveal, compare and discuss general methods and tendencies, describing all the included papers in detail is out of the scope of this work. Rather, a full list of papers is provided in Appendix A as a resource for future in-depth analysis. We further refer to Appendix A for the codes and data extraction used to synthesise the results.

\subsection{Overview of Papers}
\begin{figure*}[t]  
\centering
\fbox{\begin{minipage}{\dimexpr \textwidth-2\fboxsep-2\fboxrule}
\includegraphics[width=1.0\textwidth]{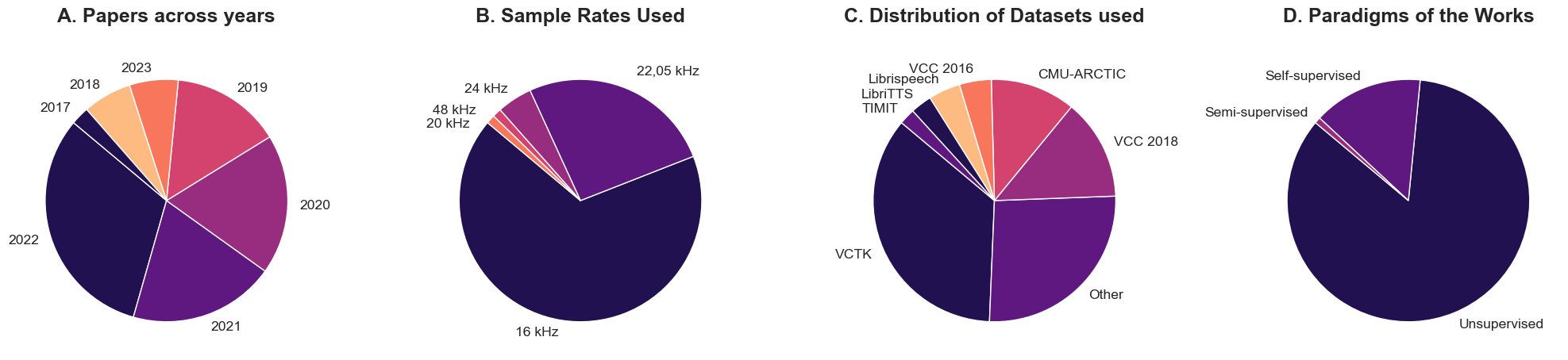}
\caption{\textit{Illustrations of the papers across years, datasets used and sample rates employed.}}
\label{fig:pies}
\end{minipage}}
\end{figure*}

The final corpus, consisting of 123 unique papers, was published in 38 different publication venues, with INTERSPEECH (40), ICASSP (21), APSIPA (8), IEEE-ACM Transactions on Audio, Speech and Language Processing (5) and ISCSLP (4) being the most popular platforms. In Figure \ref{fig:pies} we see that the work on voice conversion gradually has increased since 2017, topping in 2022 with 39 papers focusing on the topics we have screened for. This is strong evidence that interest in voice conversion still is blooming. Most of the work concentrates on systems using rather low dimensional data i.e. input sampled at 16 kHz or 22,05 kHz, which is suitable for speech since the human vocal range does not exceed the Nyquist-frequency for these sampling rates. However, this is not sufficient for modern musical applications where such sampling rates are considered low-quality. Consequently, only 1 study incorporates a high sampling rate i.e. 48 kHz. It is additionally noticed that there is consensus on training the voice conversion systems on the VCTK dataset \cite{yamagishi2019vctk} that provides speech data uttered by 110 English speakers with various accents (n=50). However, datasets such as the VCC 2018 (n=19) and the CMU-ARCTIC dataset (n=16) are other popular choices. Finally, the majority of voice conversion pipelines are constructed following the unsupervised paradigm. This alignment is consistent with the observation from related work stating that most voice conversion systems incorporate either AE's or GAN's as outlined in Section \ref{sec:background}. 

\subsection{An Overview of The Structures employed}\label{sec:structures}
\begin{figure*}[t]
\centering
\setlength{\fboxrule}{0.0001pt}
\setlength{\fboxsep}{5pt}
\fbox{\begin{minipage}{\dimexpr \textwidth-2\fboxsep-2\fboxrule}
\includegraphics[width=0.98\textwidth]{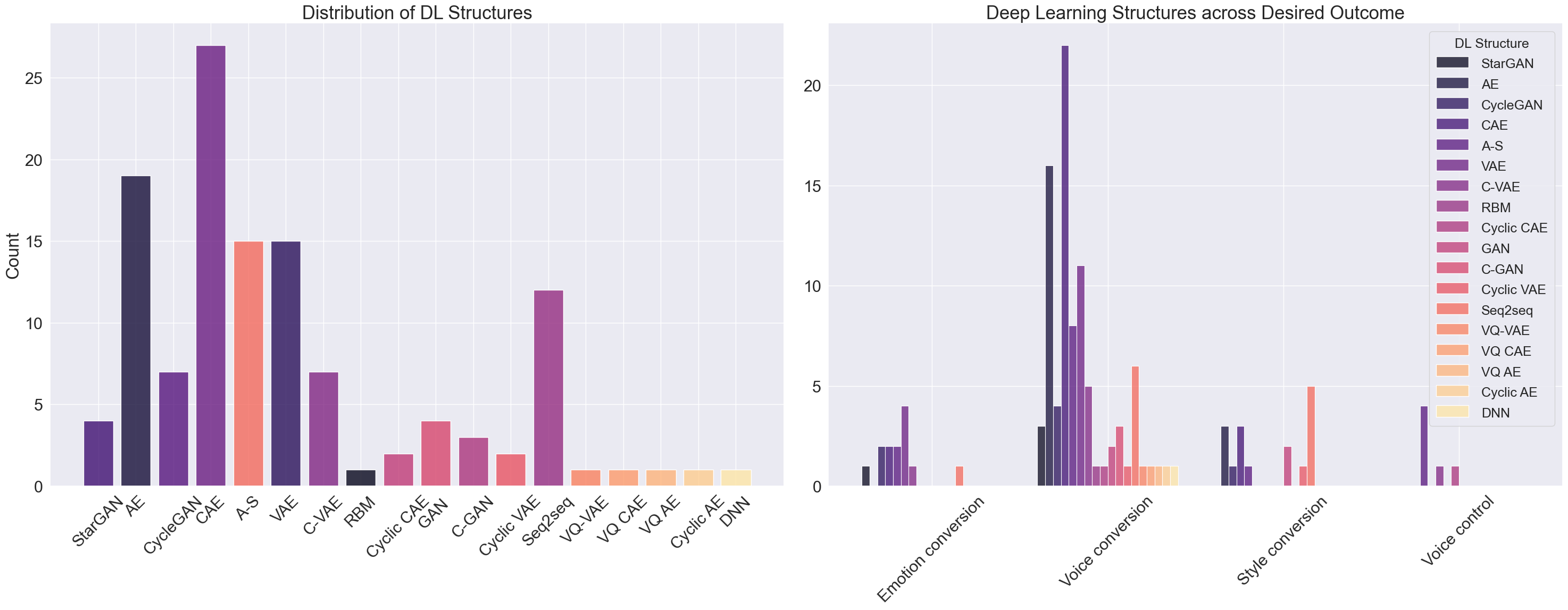}
\caption{\textit{Deep learning structures used across the different papers (left) and deep learning structures distributed across their use for specific goals and outcomes (right). "Style conversion" is a combination of both accent, prosody and cross-lingual voice conversion used for brevity.}}
\label{fig:structure}
\end{minipage}}
\end{figure*}

A main aspect of the VC pipeline is the overall structure used to learn the conversion. As shown in Figure \ref{fig:structure}, most of the structures used in the studies we have analysed are in the realm of AEs, with the conditional auto-encoder (CAE) (n=27), conventional AE (n=19) and variational auto-encoder (VAE) (n=15) being the most frequently used structures. We distinguish between CAEs and conventional AEs in their use of external conditioning features.

Among many works, they in \cite{hwang_stylevc_2022, speech-split1} and \cite{kim_assem-vc_2021} explicitly condition the decoder on both speaker and pitch embeddings in order to inform appropriate content and style representations. The conditioning can be used to tune the bottleneck as well as constrain the information flow of the speech component forcing it to be disentangled on the AE input. Given the potential entanglement of speaker style, prosody, and linguistics in the latent space, the conditioning features serve the purpose of supplying additional information to the decoder. This, in turn, guides the encoder to focus on learning only the essential and speaker-independent representation, thereby disentangling the intertwined aspects of the input data. The non-conditioned AEs and VAEs often differ from the CAEs in their end goal. Most frequently they consider one-to-one or many-to-one conversion, limiting the need for external information \cite{zang_foreign_2022, cao_nonparallel_2020}. Rarely, the conditioning is substituted with vector quantization to improve the extracted content information \cite{tang2022avqvc}. In the right-side plot of Figure \ref{fig:structure} we see that AEs and VAEs mostly are employed for style and emotion conversion whereas their conditioned counterparts in particular are popular for voice and speaker identity conversion. This is most likely due to the external and speaker-dependent information needed in zero-shot and many-to-many conversions.

CAEs are closely related to the structure we in this work call the "analysis-synthesis" structure (A-S). Nonetheless, A-S structures differ in that they are not inherently obligated to encode the parsed conditioning parameters. Contrary to CAEs, where pitch and speaker information is fed through individual encoders, A-S systems may include encoders but are not limited to doing so - explicit information extracted from DSP processes often are enough. A-S structures often decompose a signal into several attributes counting content, timbre, pitch and energy, and can, as seen in \cite{wang_vqmivc_2021, nercessian_end--end_2021, choi_neural_2021} and \cite{xie_end--end_2022} concatenate the pitch and/or energy information directly with the content and speaker embeddings before being fed to the decoder. Doing so introduces constraints on the intermediate and mid-level representations. Distinctive to A-S systems are their transparent control mechanisms. Given that their parameters are explicitly accessible and employed for training the underlying decoder or generator, A-S architectures exhibit a high level of controllability and frequently yield an interpretable control space. Further elaboration on this topic will be provided in Section \ref{sec:control}.

Besides the AE-based structures we in Figure \ref{fig:structure} see a small representation of voice conversion systems based on StarGAN (n=4) and CycleGAN (n=7). Both methods come from the field of image-to-image translation where a cycle consistency loss enables training without the need for paired data. GAN-based systems directly manipulate the input data in an attempt to generate data that closely conforms to the distribution of the target speaker. Compared to AEs this approach does not necessarily disentangle speaker information from linguistic information. Rather, it relies on the discriminator’s ability to capture the human perception of speaker identity and the generators ability to create an output that can deceive the discriminator. As GANs do not force the latent representations to be disentangled, they can preserve linguistic information and produce more natural speech. However, while the optimization of the discriminator and the adversarial losses may yield an output that resembles the distinctive characteristics of the target speaker, it does not always guarantee that the contextual information of the source speaker is kept intact. To take account of the missing contextual information the work in \cite{chun_non-parallel_2023, kaneko_cyclegan-vc2_2019, liang_pyramid_2022} introduces cycle consistency loss. In cycle consistency loss a supplementary generator carries out an inverse mapping of the target \textit{y} to the input \textit{x} during training. This procedure encourages the two generators to find \textit{(x, y)} pairs with the same contextual information, forcing the transformed output to match both the linguistics and the timbre of the target. Contrary, StarGAN-based VC systems compress the CycleGAN structure into one generalizable generator-discriminator pair. Using spatially replicated domain codes, often in the style of one-hot vectors, StarGAN conditions the system on speaker information. This allows the model to learn more than one speaker-configuration \cite{kaneko_stargan-vc2_2019, baas_stargan-zsvc_2020}. While the generators employed in voice conversion architectures based on StarGAN and CycleGAN frequently exhibit AE-like characteristics, their dimensionality reduction and lack of external conditioning make them highly non-interpretable. To address this challenge, one may combine CAEs and GANs into "adversarial auto-encoders". These are built similarly to the traditional CAE structure, but are guided by adversarial losses and have become rather popular choices for VC. This is evident as 33 of the 81 AE-based structures analysed in our review (CAE, AE, VAE, A-S), included adversarial losses.

Lastly, we note that 12 of the studies retrieved follow a sequence-to-sequence (seq2seq) procedure. While most seq2seq models are based on encoder–decoder structures \cite{DBLP:journals/corr/SutskeverVL14}, we distinguish between these models and traditional AEs as they differ significantly in their processing and training steps. Following seq2seq modelling, the work in \cite{huang_any--one_2020} encodes raw input speech into discretized features, which are represented as indices. A target-dependent seq2seq model then learns a mapping between the source feature sequence and a target sequence. Moreover, \cite{zhang_non-parallel_2020} extract a content embedding using a seq2seq recognition encoder. The encoder is here guided by embedding coming from a text encoder that is fed with phoneme transcriptions. The seq2seq structure is here an external module "adopted for obtaining disentangled linguistic representations" \cite{zhang_non-parallel_2020} and is one of many use cases of seq2seq-based models. These models are often pre-trained on big, multi-speaker datasets, and though they create a robust many-to-one pipeline, they are limited by sequential modelling and complex intermediate structures.

\begin{figure*}[t]  
\centering
\setlength{\fboxrule}{0.0001pt}
\setlength{\fboxsep}{5pt}
\fbox{\begin{minipage}{\dimexpr \textwidth-2\fboxsep-2\fboxrule}
\includegraphics[width=0.98\textwidth]{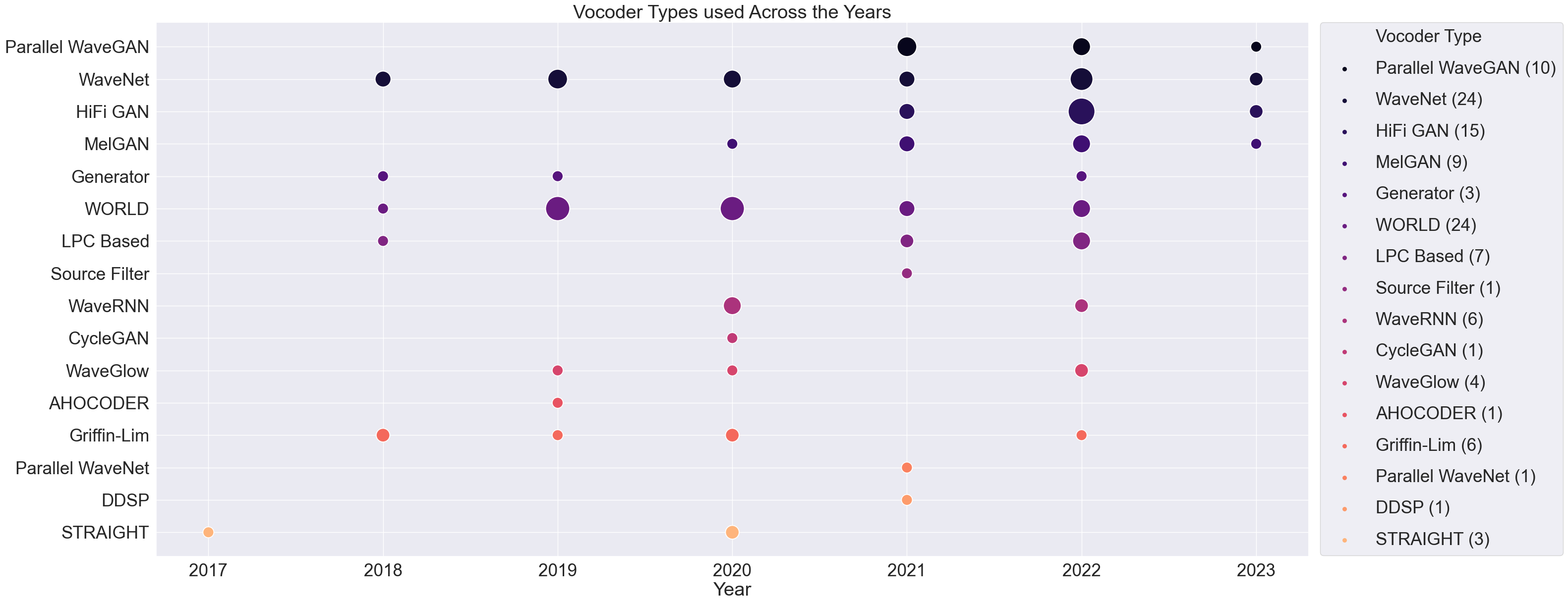}
\caption{\textit{The distribution of vocoders used from 2017-2023. The size of the shape represents the amount of similar vocoders used in that year.}
}
\label{fig:vocoders}
\end{minipage}}
\end{figure*}

\subsection{The Use of Vocoders} \label{sec:vocoders}
Vocoders are crucial to the voice conversion process, as they enable the generation of audio based on the intermediate representation. Just like the difference in deep learning structures employed, the choice of the vocoder differs depending on the use case. In general, we can divide vocoders into three main classes; 1) concatenate, signal-based models such as the harmonic plus noise model \cite{hpn}, 2) hand-designed vocoders, or source-filter models, such as the STRAIGHT \cite{STRAIGHT} and WORLD \cite{WORLD} models and 3) neural vocoders such as the WaveNet \cite{oord2016wavenet} or the WaveRNN \cite{wavernn} models.

Neural vocoders have grown increasingly popular as their data-driven approach and high-quality output allows for very expressive synthesis. They simultaneously only require an intermediate acoustic representation as input, often in the form of a mel-spectrogram, and are therefore highly flexible allowing them to be inserted in almost any end-to-end pipeline. The statistics of the analysed work in our review support these arguments. Looking at Figure \ref{fig:vocoders} it is seen that 69 of the 123 papers utilised a neural vocoder as its synthesis back-end, with the WaveNet (n=24) vocoder being the most popular.

In general, neural vocoders have become state-of-the-art for audio quality. WaveNet \cite{oord2016wavenet} paved the way in 2016 with its auto-regressive nature and is still used today. In \cite{yang_speech_2022} and \cite{bonnici_timbre_2021} it is used as the main synthesis block in AE-based pipelines whereas it in \cite{zhang_non-parallel_2020} is used in a seq2seq modelling pipeline. Like most neural vocoders, the WaveNet is conditioned via acoustic features, however, both \cite{tan_zero-shot_2021} and \cite{Wu_2021} extend it with fundamental frequency (F0) conditioning in order to force decoupling between the pitch and content. In \cite{tan_zero-shot_2021} this is done by substituting the predicted mel-spectrogram with simple acoustic features (SAF) such as the mel-cepstral coefficients (MCCs) and log-F0 information. Differently, they in \cite{Wu_2021} extend the WaveNet implementation itself through pitch-dependent dilated convolution neural networks (PDCNN) and auxiliary F0 conditioning.

Still today, WaveNet's sequential generation is prohibitively costly motivating the development of more efficient neural vocoders. As a result of this, GAN-based vocoders have become a new benchmark for neural vocoders due to their fast inference speed and lightweight networks \cite{overviewSisman}. In the analysed studies the most widely used GAN-based vocoders are the HiFi-GAN (n=15) and the Parallel WaveGAN (n=10). In Figure \ref{fig:vocoders} we see that their use has become more frequent after the invention of the MelGAN in 2019. Besides the inclusion of adversarial training, the inputs to and usage of GAN-based vocoders do not differ significantly from other neural vocoders such as the WaveNet and the papers examined rarely justify the type of vocoder chosen. However, it is often mentioned that GAN-based vocoders are included due to their "better speech quality and much faster inference speed" \cite{lian_towards_2022}.

Even though neural vocoders have become increasingly popular, parametric and hand-designed vocoders still are used. Figure \ref{fig:vocoders} shows that the WORLD vocoder was used more than its neural counterparts in the years 2019-2020 (n=24). The WORLD vocoder is a high-quality speech synthesis and analysis tool used for extracting and synthesizing waveform information. It is in the work analyzed mainly used for two reasons; firstly, its inherent capability to extract pitch and timbre information provides a strong foundation for subsequent disentanglement efforts; secondly, the substantial amount of acoustic data it offers facilitates a straightforward guidance of a WORLD synthesis process. The work in \cite{huang_unsupervised_2020} and \cite{kaneko_stargan-vc2_2019} as an example use the WORLD vocoder to extract aperiodicity signals (APs), F0 features and 513-dimensional spectral envelopes. The spectral envelopes are further reduced to more specific MCCs and encoded for linguistic/content information. The F0 features are linearly transformed to match the target, and can, together with the extracted AP information, be carried over to the inverse WORLD synthesis stage directly. This simplifies the conversion task to be a non-linear transformation of the source spectrum only (often conditioned on extra target speaker information). Almost the same procedure can be carried out for the STRAIGHT vocoder, however, this vocoder has only been utilised in 3 of the analyzed works. Though these parametric synthesizers offer robustness and flexibility, they are limited to monophonic reproduction. They are simultaneously limited by their internal synthesis mechanism, which often produces artefacts \cite{nguyen_nvc-net_2021}.

Closely related to the hand-designed vocoders are the signal-based vocoders used in 2 of the studied works. An example of this is the continuous sinusoidal model utilised in the synthesis stage of \cite{al-radhi_effects_2021}. Here a neural network converts sinusoidal parameters, constructing speech frames from a voiced and a noise component respectively. Since the synthesis stage is vocoder-free, this approach simplifies the learning process limiting the model to learn the reconstruction of the intermediate representation only. Similarly, a harmonic plus noise model is utilised in \cite{nercessian_end--end_2021}. Inspired by the differentiable digital signal processing approach (DDSP) \cite{engel2020ddsp}, a feature transformation network learns to map input attributes to the parameters for a differentiable H+N synthesizer. More specifically the network predicts the harmonic distribution for an additive sinusoidal synthesiser and 65 noise filter taps used to filter the noise part of the produced speech signal. While being both efficient and lightweight, the inclusion of the DDSP framework additionally introduces an inductive bias as the sinusoids may be directly controlled by the input pitch. Contrary, the output quality is limited by the capabilities of H+N synthesis, forcing post-filtering or extraneous processing. 

While most of the work examined in this review uses actual vocoders and thus adheres to an encoder-decoder-vocoder structure, few works are taking a more immediate approach using the generator to produce time-domain data (n=3). With the reasoning that traditional VC highly depends on both the quality of the intermediate representation and the vocoder itself, the NVC-Net in \cite{nguyen_nvc-net_2021} as an example perform "voice conversion directly on the raw audio waveform". This is done by combining the decoder and the vocoder into a single generator inspired by the MelGAN \cite{kumar2019melgan}. More specifically, the generator upsamples the content embedding using transposed residual convolutions with each residual connection being conditioned by a target speaker identity. By limiting the NVC-Net to exploit only its internal representation, the authors provide a high-quality, condensed and fast framework claimed to "generate samples at a rate of 3661.65 kHz on an NVIDIA V100 GPU in full precision and 7.49 kHz on a single CPU core" \cite{nguyen_nvc-net_2021}. These methods are therefore highly useful in low-latency scenarios, but also favourable for zero-shot VC as the speech production is independent of intermediate representations and mismatch problems.

\subsection{Choices of Feature Extraction}
Choosing the encoder structures for both feature extraction and the feature embeddings is another important aspect of the VC pipeline. As earlier mentioned it is crucial to isolate both the content and the speaker identity in order to disentangle linguistics from speaker characteristics.

\subsubsection{Content Embeddings}
\begin{figure*}[!t]  
\centering
\setlength{\fboxrule}{0.0001pt}
\setlength{\fboxsep}{5pt}
\fbox{\begin{minipage}{\dimexpr \textwidth-2\fboxsep-2\fboxrule}
\includegraphics[width=0.98\textwidth]{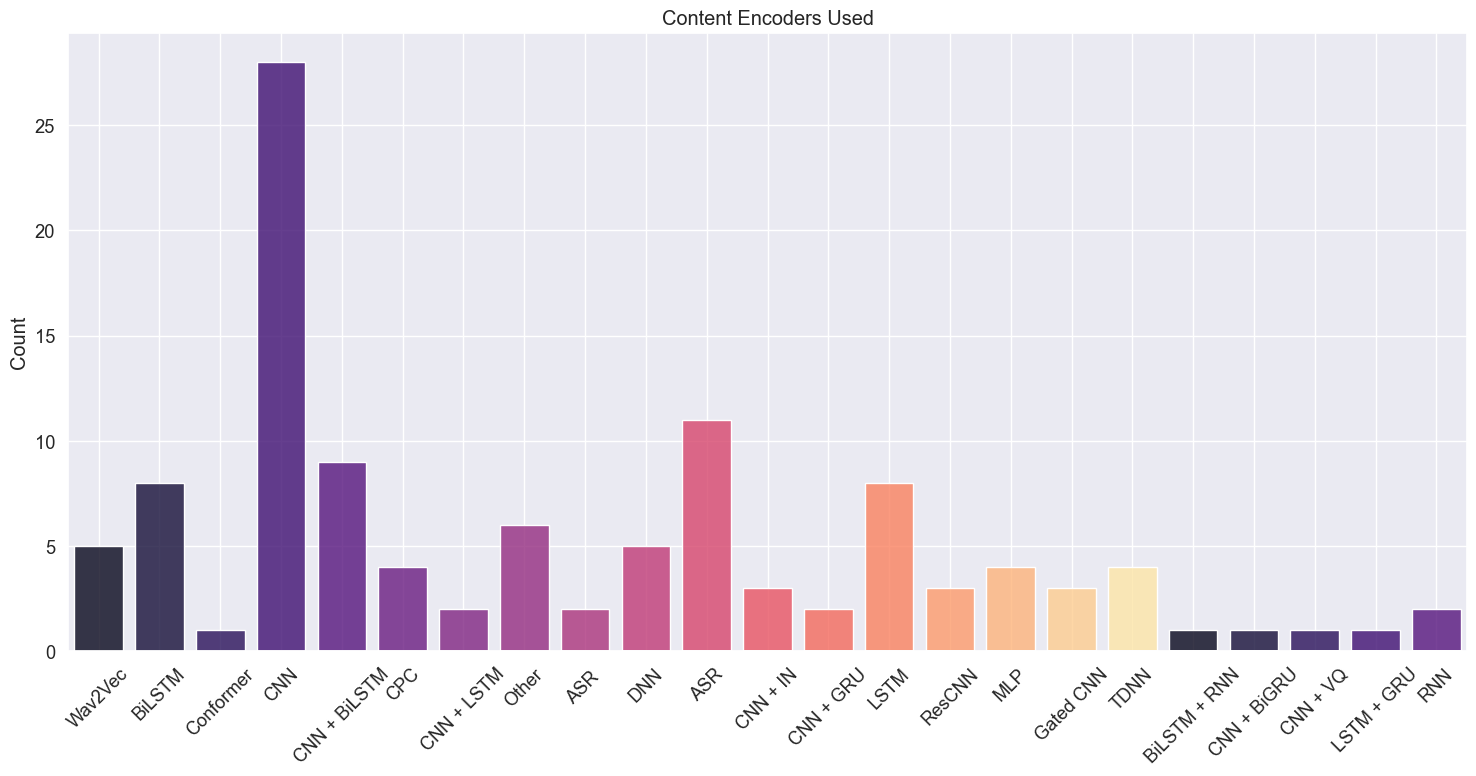}
\caption{\textit{The distribution of content encoder networks used across all the analysed studies. For brevity, we have collected encoders appearing once in the 'other' category. This category, among others, contains pyramid attention modules, feed-forward transformers and dynamic time-warping models.}
}
\label{fig:content}
\end{minipage}}
\end{figure*}

Looking at Figure \ref{fig:content} there is consensus on the networks used for retrieving linguistic content information, with traditional CNNs and related structures being the most popular. In the analysed work, singular CNNs are used to extract the content embedding in 28 of the cases, whereas the combination of CNNs and BiLSTMs are incorporated 9 times. The combination of CNNs and recurrent neural networks (RNNs) such as the BiLSTM is interesting as it is inspired by the field of automatic speech recognition (ASR). In ASR systems, CNNs are employed as they are beneficial in modeling local acoustic patterns, either in the audio signal or the spectrogram, while the RNN is advantageous in capturing temporal dependencies. Overall, the analyzed work represents ASR-based content information as; 1. linguistic embeddings using condensed ASR blocks such as the CNN + BiLSTM or CNN + LSTM encoders trained end-to-end in \cite{tan_zero-shot_2021, choi_sequence--sequence_2021, wang_drvc_2022}, or 2. as PPGs obtained from speaker-independent ASR systems as done in \cite{mohammadi_one-shot_2019} and \cite{chen_improving_2022}. The latter is efficient as it utilises pre-trained models for the extraction task, often trained on large, multi-lingual datasets such as the 'kaldi speech recognition toolkit' \cite{kaldi}, the 'julius dictation kit'\footnote{https://github.com/julius-speech/dictation-kit} or through the conformer model \cite{gulati2020conformer}. Obtaining linguistic content through speech recognition models is useful as it
decodes linguistic discriminant information from speech without the consideration of who is speaking, thus creating a complete speaker-independent representation. Overall, the ASR-based approach "frees up the conversion network from using its capacity to represent low-level detail and general information" \cite{overviewSisman}. Rather the network has the possibility to focus on the high-level semantics necessary for speaker identity conversion. 

However, the PPG based approach may still include errors from the recognition model itself leading to mispronunciation in the converted speech output. As pictured in figure \ref{fig:content} we therefore additionally experienced a small representation of wav2vec models used to extract the linguistic content information (n=5). Wav2vec models can be either unsupervised \cite{schneider2019wav2vec} or self-supervised and are furthermore often employed to output speech representations in ASR pipelines \cite{baevski2020wav2vec}. Similar to the aforementioned ASR systems, wav2vec models provide linguistic embeddings that are time-aligned and speaker-independent. Their high-level analysis features make them superior on downstream tasks, especially for low-resource languages \cite{choi_neural_2021}. In \cite{chun_non-parallel_2023} and \cite{choi_neural_2021}, the content embedding is derived from the 12th layer of a pre-trained XLSR-53 wav2vec model. This particular representation has been chosen due to its reported significance in encapsulating essential pronunciation-related characteristics \cite{shah2021audio}. 
\begin{table*}[!t]
\renewcommand{\arraystretch}{1.3} 
\fbox{\begin{minipage}{\dimexpr \linewidth-2\fboxsep-2\fboxrule}
\caption{\textit{Overview of analyzed VC pipelines that are focusing on disentangled speech representation learning using 4 or more explicit features. 'Main goals' describe the desired outcome of the work, 'pitch and feature representation' describe the methods used to represent the given feature, 'main contribution' describe the primary technique used to achieve the end goal}}
\label{tab:features} 
\vspace*{1ex}
\scriptsize
\begin{tabularx}{\linewidth}
{p{0.13\linewidth}|X|p{0.20\linewidth}|p{0.20\linewidth}|p{0.20\linewidth}}
\hline
\textbf{Reference} & \textbf{End Goal} & \textbf{Pitch Representation} & \textbf{Feature Representation} & \textbf{Main Contribution}\\
\hline
S. Yang et al \cite{yang_speech_2022} & One-Shot VC & RAPT $\rightarrow$ Z-Norm  $\rightarrow$ Enc & \textit{Rhythm}: Enc (BiLSTM) & Mutual information learning \\
J. Wang et. al \cite{wang_adversarially_2021} & Any-to-many VC & Contour $\rightarrow$ RS $\rightarrow$ Enc & \textit{Rhythm}: Enc (CNN, BiLSTM) & Adversarial MAP \\
Z. Luo et. al \cite{luo_decoupling_2023} & Emotional VC & Contour $\rightarrow$ RS $\rightarrow$ Enc & \textit{Rhythm}: Enc (CNN, BiLSTM) & Source-Filter based \\
\hline
Y. Wang et. al \cite{wang_controllable_2022} & Voice Control & Swipe + CREPE $\rightarrow$ Abs-Norm & \textit{Energy:} Time-domain & Adv. training \& AIC Loss \\
H. Choi et. al \cite{choi_neural_2021} & Voice Control & Yingram (Yin spectrogram) & \textit{Energy:} Avg. Log-Mel Spec & Information Perturbation \\
S. Wang et. al \cite{wang_zero-shot_2022} & Zero-shot VC & Enc (RankNet) & \textit{Energy:} Enc (RankNet) & Self-supervision \\
S. Nercessian \cite{nercessian_improved_2020} & Zero-shot VC & CREPE $\rightarrow$ Log & \textit{Energy:} A-weighted spec & Explicit conditioning \\
Q. Xie et. al \cite{xie_end--end_2022} & Any-to-any VC & Contour $\rightarrow$ Enc & \textit{Energy:} Time-domain & Information Perturbation \\
\end{tabularx}
\end{minipage}} 
\end{table*}

The input to the content encoder may vary depending on the feature extraction method and the vocoder utilized. Nevertheless, it is important to emphasize that in nearly all cases, the input primarily consists of acoustic features such as the mel-spectrogram or MCCs.

\subsubsection{Speaker Embeddings}
Contrary to the content encoders, the techniques used to retrieve speaker embeddings and timbre characteristics differ significantly across the field. Few works obtain speaker embeddings by averaging the frame-level characteristics of different speaker utterances or by downsampling the input to one-hot vectors \cite{kaneko_stargan-vc2_2019}, while others incorporate feature vectors and codebooks directly \cite{reddy_dnn-based_2020, ho_cross-lingual_2021}. Few works use the ECAPA-TDNN architecture trained for speaker verification \cite{zhang_sig-vc_2023} or the earlier mentioned XLSR-53 model whose 1st layer forms clusters for each speaker \cite{choi_neural_2021}. The authors in \cite{du_high_2023} further extend the ECAPA-TDNN by pre-trained X-vector networks to address the differences in their distributional variations. More simple and traditional approaches to speaker embeddings are also taken. In \cite{dang_training_2022} a 12-layer CNN is used to encode the input mel-spectrogram, while they in \cite{chen_improving_2022} use a BiLSTM-based speaker encoder pre-trained for speaker classification. It is essential to acknowledge that many of the less intricate algorithms frequently are supported by adversarial, classification, or cycle-consistency losses. We will this discuss this in Section \ref{sec:loss}.

\subsubsection{Additional Embeddings}
Besides the mentioned content and speaker information, the works analysed include the following features for conditioning and further embedding; pitch (n=49), prosody/rhythm (n=8), energy (n=6) and emotion (n=5). As earlier mentioned, pitch extraction is commonly applied to the VC pipeline. When embedding linguistics and speaker characteristics, speaker identity may in many cases be disentangled from speech content. However, a significant amount of prosodic information, such as volume or source F0, is often still entangled in the content embedding, which in turn may leak into the intermediate representation causing mismatch problems or making the converted F0 fluctuate. In this context, the inclusion of supplementary pitch-related data can prove advantageous, serving a dual purpose; firstly it enables the content encoder to concentrate exclusively on linguistic aspects, secondly, as demonstrated when integrating the WORLD vocoder, it introduces an inductive bias by directly incorporating pitch information into the synthesis phase. 

In Table \ref{tab:features} we provide an overview of the VC pipelines that include 4 or more different features and note the techniques used to extract the features outside of the traditional content and speaker conventions. We additionally summarise the overall goal of doing so. As seen in \ref{tab:features} the studies listed mostly take a more explicit approach to disentangled speech representation learning, using both pitch and prosody information. It is evident that the rhythm embedding often is carried out by encoder structures similar to the ones utilised for content and speaker/timbre representations. Pitch, however, may be retrieved in terms of pitch-contour using classical signal processing methods like the YIN \cite{yin}, RAPT \cite{rapt} or CREPE \cite{crepe} algorithms. In all analysed cases the pitch contour is further processed by a pitch encoder. In \cite{xie_end--end_2022} this allows the pitch embedding to be influenced by information from the target speaker, whereas it in \cite{wang_adversarially_2021} and \cite{nercessian_improved_2020} is used to create a more condensed representation. The encoding of energy information is contingent upon the decoder structures employed. Nonetheless, it is most commonly conveyed as an additional, unaltered conditioning, owing to its classification as less critical data.

\subsection{Losses, costs and errors}
\label{sec:loss}
As outlined above, the primary determinant of the success in voice conversion lies in the specific deep learning structures employed, such as the information bottleneck principle of the underlying AE system or the generative capabilities of the GAN. However, their outcome can be improved, and intermediate tasks may be supported by loss functions specifically tailored to optimize them. An obvious choice is to train the system using a 'reconstruction loss', which in the analyzed studies was mentioned 74 times. Most of the work aims at minimizing the reconstruction loss in the acoustic feature space, that is, the difference in spectral envelopes \cite{he_improved_2021}, mel-cepstral coefficients \cite{kaneko_cyclegan-vc_2018} or the difference between the input mel-spectrogram $X_1$ and the predicted mel-spectrogram $\hat{X}_{1\rightarrow1}$, before vocoding. The reconstruction loss can be generalized by the equation:

\begin{equation}
    L_{recon} = \mathbb{E} [ \|\hat{X}_{1\rightarrow1} - X_1\|],
\end{equation}

where the difference between the prediction and the ground truth may be calculated using the L1 or the L2 distance, the mean squared error (MSE) or the mean absolute error (MAE). Rather than minimizing the error between the acoustic features, few works calculate the reconstruction loss in the time domain \cite{du_high_2023}. Some studies additionally extract a perceptually based spectral loss from the produced waveforms \cite{choi_neural_2021}. The spectral loss compares the input spectrogram with the spectrogram of the time-domain output and helps the generator obtain the time-frequency characteristics of the produced speech. The predicted and ground truth spectrograms may be compared individually \cite{nguyen_nvc-net_2021} or at multiple resolutions as done in \cite{nercessian_end--end_2021}. Both in the case of the time domain and the perceptual reconstruction losses, the vocoder is incorporated into the learning process giving more degrees of freedom.

As an alternative approach to the reconstruction loss, the VC system may be trained using a feature-matching loss (FM loss). The FM loss is an adversarial method that incorporates a discriminator to "guide" the reconstruction. The global network is then trained in conjunction with the discriminator, while the generator is updated based on the similarity between the prediction and the ground truth feature maps produced by the discriminator:

\begin{equation}
    L_{FM} = \mathbb{E}_{(c,z,x)} \Bigl[ \sum_{i=1}^{L} \frac{1}{N_i} \| D_{i} (x) - D_{i} (G(c,z) \|_1 \Bigr]
\end{equation}

Here $D_i$ denotes the feature map output of $N_i$ units from the discriminator at the i-th layer \cite{nguyen_nvc-net_2021}. A total of 8 studies incorporate FM-loss and it is used both for GAN-based pipelines \cite{lu_towards_2021} as well as adversarially tuned CAEs and C-VAEs \cite{xie_end--end_2022, huang_flowcpcvc_2022}. Besides FM-loss, the reconstruction loss may be extended by a content loss. The content loss is particularly useful for constraining the information capacity of the bottleneck, as such systems often are invariant to self-to-self reconstruction. Content loss is among others applied in the works of \cite{wang_controllable_2022, li_asgan-vc_2022} and \cite{shi_u-gat-vc_2022}, and given by the equation \cite{qian2019autovc}:

\begin{equation}
    L_{content} = \mathbb{E} [ \|E_c(\hat{X}_{1\rightarrow1}) - C_1\|],
\end{equation}

where $E_c(\hat{X}_{1\rightarrow1})$ is the content embedding of the prediction and $C_1$ the actual content embedding of the input e.g. $E_c(X_1)$. Incorporating feature specific loss functions like the content loss is a commonly employed approach. This is evident as 10 of the papers we examined utilise feature specific loss functions besides the ones based on content. In \cite{yang_speech_2022}, a pitch-contour loss is used to compare the pitch of the analysed input with the pitch of the reconstruction to enforce pitch coherence. Similar to the content loss, in \cite{lee_many--many_2021} the style embedding of the input is compared with the prediction fed through the style encoder. Lastly, studies optimize speaker embeddings either by the inclusion of classification losses (n=9), cross-entropy losses (n=9) or speaker-ID losses (n=4). In \cite{ho_cross-lingual_2021} and \cite{ding_study_2022} an auxiliary classification model is trained to help the generator "produce fake data with the correct target speaker voice". The classifier is trained using the output log-likelihood that the acoustic features coming from the generated audio belong to the target speaker. In \cite{chen_improving_2022} and \cite{dang_training_2022}, on the other hand, speaker similarity is improved by optimizing the speaker encoder with a speaker classification loss based on the ground-truth speaker identity label in one-hot vector format. Based on this, we define the general speaker identity loss:

\begin{equation}
    L_{speaker} = CE\bigl(x_{id}, softmax(V * E_s(x))\bigr),
\end{equation}

where CE is the cross-entropy loss, $x_{id}$ the ground-truth speaker identity vector, $V$ a trainable weight matrix and $E_s$ the speaker encoder \cite{chen_improving_2022}. The speaker ID also plays a role in what is called the "cycle-consistency loss" discussed next.

Adversarial and cycle consistency loss functions are incorporated in 36 and 21 of the studies respectively. As earlier mentioned, the use of adversarial losses is not limited to GANs. In \cite{huang_winvc_2021} a very traditional adversarial loss guided by a patchGAN discriminator is used in an AE pipeline. In \cite{xie_end--end_2022} an adversarial loss based on parallelWaveGAN is incorporated in conjunction with a multi-period discriminator (MPD), multi-scale discriminator (MSD) and multi-resolution spectrogram discriminator (MRSD) to steer the A-S process. Here the model parameters are optimized based on the generator's ability to deceive all discriminators. Lastly, in \cite{hwang_stylevc_2022} the discriminative process is extended by a pitch-based discriminator that in addition to the real/fake probability prediction, predicts how much the pitch of the reconstruction is similar to that of the target speaker. Most of the cycle-consistency losses included take a form based on the StarGAN paradigm; rather than incorporating another generator they use the main generator to map the prediction back to its original form by including a speaker label (often referred to as a domain classifier). This is among others done in \cite{zhang_gazev_2020} and \cite{huang_winvc_2021} following the form:

\begin{equation}
    L_{cyc} = \mathbb{E}_{c\sim(c),x\sim(x|c'),c\sim(c)} \big[ \|G(G(x,c),c') - x \|\big],
\end{equation}

with $c$ representing the attribute label that classifies the domain and $G$ being the generator. Cycle consistency loss aims to enhance the contextual robustness of the encoder. When the adversarial loss forces output to follow the target-data distribution, the cycle-consistency loss is utilised to preserve the composition in the conversion.

While cycle consistency provides a constraint and encourages the forward and inverse mappings to find (x, y) pairs with the same contextual information, it does however not guarantee that the mappings always preserve linguistic information. In order to do so, the identity mapping loss is included. We find identity mapping losses in 13 of the works. Identity mapping losses are usual in cycleGAN-based pipelines and are, equally to the content loss, used to preserve linguistic information without relying on extra modules \cite{kaneko_cyclegan-vc_2018}. In general, the identity mapping loss is adopted to regularize the generator to be close to an identity mapping when one converts the input to that of the same speaker. As mentioned in \cite{cao_nonparallel_2020} the intuition behind this is that "the model is supposed to preserve the input if it already looks like that from the target domain". We can represent the identity-mapping loss by the equation:
\begin{equation}
    L_{id} = \mathbb{E}_{y\sim p_y(y)} \big[ \|G_{X\rightarrow Y} (y) - y \|_1\big] +  \mathbb{E}_{x\sim p_x(x)} \big[\|G_{Y\rightarrow X} (x) - x \|_1\big]
\end{equation}

In this context, let $G_{X\rightarrow Y}$ represent the generator responsible for the transformation from the source domain to the target domain, and $G_{Y\rightarrow X}$ denote an auxiliary generator tasked with performing the reverse transformation. The role of the auxiliary generator $G_{Y\rightarrow X}$ is to facilitate the preservation of composition between the input and output domains. This encourages the primary generator, $G_{X\rightarrow Y}$, to discover the mapping that effectively maintains the compositional integrity throughout the transformation process.

\begin{figure}[t]  
\centering
\setlength{\fboxrule}{0.0001pt}
\setlength{\fboxsep}{5pt}
\fbox{\begin{minipage}{\dimexpr \textwidth-2\fboxsep-2\fboxrule}
\includegraphics[width=\textwidth]{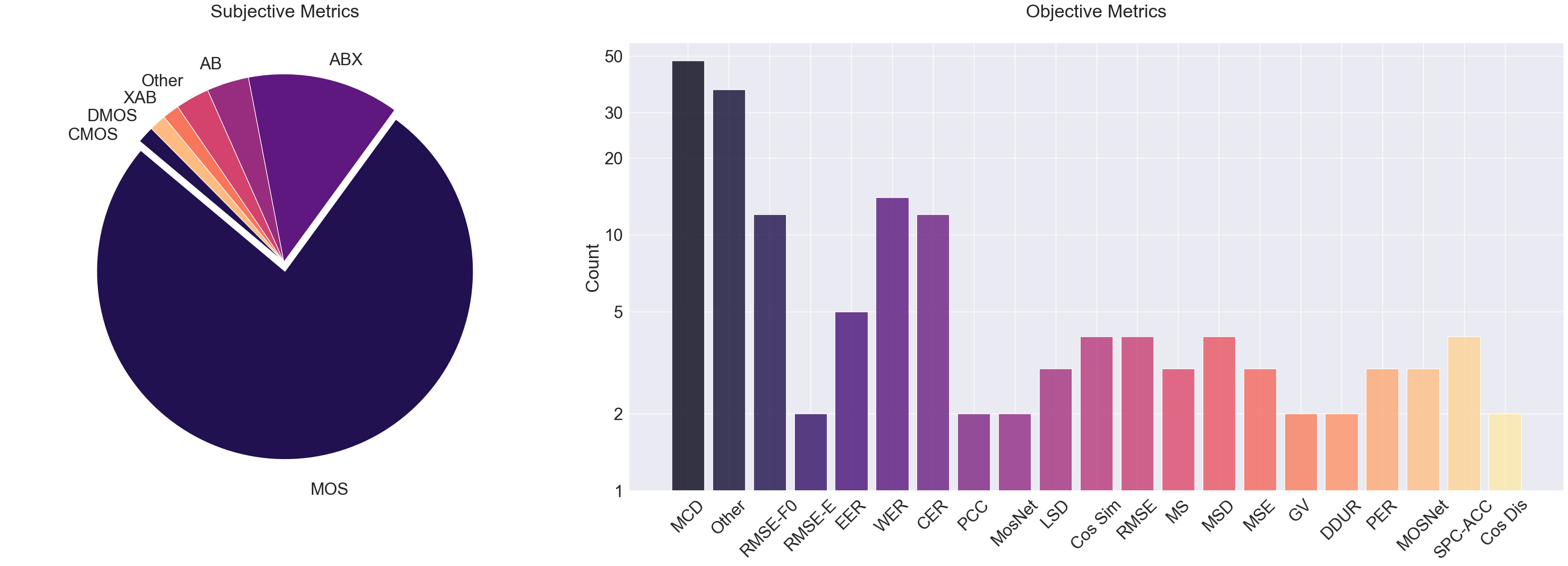}
\caption{\textit{Distribution of subjective metrics used (left) and objective metrics employed (right) across the studied work. We see that the MCD is a commonly used objective metric, while the word error rate and character error rate also are popular choices.}}
\label{fig:metric}
\end{minipage}}
\end{figure}
\subsection{Metrics and Performance Evaluation}
With the amount of VC work available, effective evaluation of the different results is required. This is needed to validate the voice quality of the system proposed, but also to compare and benchmark results against state-of-the-art work and related techniques. Most commonly, the works analysed utilise the mel-cepstral distortion (MCD) metric (n=48) to objectively evaluate the overall audio quality of the system output, compared to a reference speech sample. MCD is a measure of the difference between two sequences of mel-cepstra and though it is not always correlated with human
opinions it evaluates perceptually relevant features (MCEPs) of the two signals. We in Figure \ref{fig:metric} further see that different versions of the root mean squared error (RMSE) are used to examine specific attributes of the system outputs. RMSE-F0, RMSE-Energy and general RMSE metrics are respectively used in 12, 4 and 3 of the studies, highlighting the fact that the performance in reconstructing prosodic information is of high importance in VC systems.

Secondly, we see a high use of metrics borrowed from the speech recognition or machine translation field. The word error rate (WER) metric, which measures the percentage of words that are not correctly transformed from speech-to-text (STT), is included in 14 of the works analysed. Similar tendencies are experienced for the character error rate (CER) that is included in 12 of the papers. CER measures the percentage of characters from the output that are transcribed incorrectly. For measuring both the WER and the CER it should be noted that both the output and the reference samples are fed through an external STT engine in order to evaluate the intelligibility of the produced speech. Finally, it is observed that speaker-based metrics are employed to assess the system's capability in either transforming or retaining speaker identity. Notably, the most commonly utilized objective metrics in this regard are speaker classification (n=4) and equal error rate (EER) originating from the domain of speaker verification (n=5). 

Even though the above objective metrics are good at evaluating important aspects related to the quality of the voice conversion, they have difficulties measuring the naturalness of the reconstruction and are unable to judge its perceptual similarity to a target sample. To measure this, the VC community opt for subjective metrics such as the mean opinion score (MOS), MUSHRA scales, or AB preference tests. The MOS is a listening test asking participants to rate the conversion quality from 1 to 5, often based on the aspect of naturalness. MOS is a standard benchmark and used is in 105 of the 123 analysed studies (see Figure \ref{fig:metric}), emphasising its value in comparing perceptual aspects of the produced speech to related work. There exist evaluation methods similar to the MOS such as the degradation mean opinion score (DMOS) or the comparative mean opinion score (CMOS), which are not equally popular as the MOS (n=2). Both the DMOS and the CMOS rate the reconstruction in relation to a reference sample; the former asks the participants to rate the degradation of the reconstruction whereas the latter asks to rate the identicality. Another commonly employed subjective metric is the AB/ABX test used in 23 of the papers studied. In AB/ABX tests, participants are exposed to both the reconstructed audio and the reference audio where-after they are required to specify which one exhibits a greater degree of a specific attribute \cite{overviewSisman}.

While subjective evaluation metrics suggest a valid perceptual measure of the inherent naturalness of the VC results, it is important to acknowledge that they require a substantial number of participants to ensure validity and can be a time-consuming process. Nonetheless, they have become a standard in evaluating VC results and were in essence used in almost all studies analyzed.

\section{Discussion}
In the following Section we summarise our results and, based on our codebook, highlight 4 specific challenging areas that were discovered in the analysed studies. The challenges will be thoroughly examined, with a focus on the methodologies employed to address them in the VC field. Based on these, we furthermore explore the potential opportunities in overcoming the challenges and provide recommendations for future research.  

\subsection{Summary of Results}
Even though the global voice conversion procedure was similar for most of the studies analysed, it was found that the work highly differed in their underlying disentanglement techniques, vocoders and loss function used. In general, the work studied in this review can be divided into \textit{\textbf{AE-based}} pipelines (n=77), \textit{\textbf{GAN-based}} pipelines (n=17) and \textit{\textbf{analysis-synthesis-based}} pipelines (n=15). In general, the AE and GAN-based systems are similar in their structure using an information bottleneck, however, the GAN-based works combine the analysis and synthesis process into a single generator whereas the AE pipelines separate the encoder and decoder in order to more easily incorporate conditioning. Papers using these paradigms typically address 4 different topics; \textit{\textbf{conventional voice conversion}} counting zero-shot, many-to-many and general output quality improvements (n=88), \textit{\textbf{style conversion}} counting accent, dialect and prosody conversion (n=16), \textit{\textbf{emotion conversion}} (n=13) and \textit{\textbf{voice control}} counting works focusing on externalizing parameter control (n=6). The works primarily comprise the \textit{\textbf{WORLD vocoder}} (n=24), the \textit{\textbf{WaveNet}} (n=24), and the \textit{\textbf{HiFi-GAN}} vocoder (n=15), with the former being prevalent before the 2020s and the latter gaining popularity after its introduction in 2019. Depending on the topic addressed and vocoders used, the works additionally differ in the loss functions included. \textit{\textbf{Reconstruction losses}} (n=74) and \textit{\textbf{adversarial loss}} functions (n=36) are frequently used, but are often extended by; \textit{\textbf{cycle consistency losses}} (n=21) constraining the generator, \textit{\textbf{kullback-Leibler divergence}} (n=17) ensuring that the learned distribution is similar to the true prior distribution, and \textit{\textbf{identity-mapping losses}} (n=13) or \textit{\textbf{content losses}} (n=3) providing contextual robustness.

\subsection{Challenging Areas in the VC community}
While it is clear VC is a promising, fast-growing research area with proposed models reaching high-quality output both in terms of speaker similarity and naturalness, some challenges still need to be overcome. Drawing from these discoveries, we formulate four challenges that are commonly addressed in VC research and offer a synopsis of how the analyzed studies endeavour to resolve them.

\subsubsection{Disentangling features} \label{sec:disentangling}
Though the information bottleneck paradigm aims at disentangling speech content and speaker characteristics, a significant amount of prosodic information, such as source pitch, may leak through the bottleneck. Simultaneously, the naturalness of the converted speech may decrease due to the difference in information contained within the bottleneck features themselves. Theoretically, these problems could be solved by a speaker embedding that contains the target speaker’s prosodic information. This, however, requires hours of data for each speaker. Instead, it has been proposed to disentangle all three features; speech content, F0 and identity \cite{Qian_2020}. Among other works, this is done in \cite{huang_investigation_2019, wang_controllable_2022} and \cite{tan_zero-shot_2021} where either the generator, the vocoder, or both, are conditioned on extracted F0 information. In all cases the F0 conditioning improves the naturalness and similarity of the generated speech, however, the amount of F0 information kept in the content and speaker embeddings is highly dependent on the input representation and the amount of F0 information that resides in its spectral features. Therefore, a majority of the studies found it to be more efficient to disentangle the conditioning features from MCCs or MCEPs in contrast to the conventional and harmonically rich mel-spectrogram. To eliminate said problems of harmonically rich input representations, the authors in \cite{chun_non-parallel_2023} and \cite{hwang_stylevc_2022} utilise the wav2vec representation as input due to its linguistic-informed, but high-level layers. Moreover, in \cite{hwang_stylevc_2022} pitch is predicted from the wav2vec-based content embedding using an external predictor, while the speaker embedding is retrieved from a style encoder fed with the source mel-spectrogram. Both works report mel-cepstral distortion (MCD) and RMSE-F0 metrics that outperform those of the system explicitly conditioned on pitch. However, these works rely on non-interpretable and inefficient self-supervised networks that do not necessarily provide auxiliary pitch control.

To further disentangle the speech features, techniques such as instance normalization (IN) and vector quantization (VQ) are utilized. In \cite{zhang_gazev_2020} and \cite{huang_winvc_2021} adaptive instance normalization (AdaIN) is introduced to adjust the speaker embedding to different styles on a per-instance basis. IN and AdaIN are normal features in style transfer problems as their inherent scale and bias parameters allow the speaker modulation to be transformed in a domain-specific manner, limiting it from bleeding into the content embedding \cite{kaneko_stargan-vc2_2019}. VQ on the other hand, is incorporated to confine the leakage of content information into the speaker representation. In \cite{wang_vqmivc_2021, wu_one-shot_2020} and \cite{chen_unsupervised_2020} they as an example apply VQ on the content embedding, modelling it as a series of discrete codes. The motivation behind employing VQ lies in the observation that the discrete latent codes acquired from VQ-based auto-encoders exhibit a strong correlation with phonemes. A vector quantized content embedding simply is a more condensed representation, providing just the needed information for the decoder. However, the mapping of continuous values to a set of discrete codebook entries may lead to the loss of fine-grained information. It is therefore crucial to have a large codebook size, which in turn imposes higher complexity and memory usage. It is thus clear, that many possibilities exist in optimizing the feature disentanglement in the VC bottleneck and that methods used depend on the use case and implementation device at hand. Considering both the amount of the harmonic information in the input and the interplay between the content and speaker encoders may improve the disentanglement in the latent space.

\subsubsection{Approaching Mismatch Problems}
Most VC pipelines suffer from mismatch problems. One such is the training-inference mismatch problem happening when a VC system is trained using the same utterance from the same speaker, which is the case in most systems. Here the same input sample is used for content and style/speaker embeddings, making the overall model prone to copying information. This becomes a problem in inference as the model here is presented with different samples, making it produce low output quality. In \cite{hwang_stylevc_2022} they tackle this mismatch problem using adversarial style generalization. More specifically, the style generalization clusters representations using two different utterances of the same speaker during training. The style encoder is thus trained on both utterances and optimized by minimizing the difference between their output speaker embeddings. This creates a global style representation for each speaker that is robust in inference scenarios.

Another issue related to mismatch is the acoustic feature mismatch problem. This becomes apparent when there is a significant disparity between the acoustic representation and the generalization capabilities of the vocoder used, resulting in reduced output quality. In \cite{du_high_2023} they approach this by discarding the decoder completely, substituting it with a HiFi-GAN-based generator that up-samples the intermediate acoustic representation to match the dimensions of the time domain. Contrary, they in \cite{xie_end--end_2022} and \cite{choi_neural_2021} avoid the mismatch problem by the introduction of information perturbation. As an extension to the above-mentioned feature disentanglement methods, information perturbation aims at perturbing all useless information in the source speech through digital signal processing, thereby limiting the different sub-blocks from learning undesirable attributes \cite{xie_end--end_2022}. Specifically, they in \cite{choi_neural_2021} perturb the audio given to the content encoder with pitch shifting, formant shifting and random frequency shaping, forcing it to adhere only to the linguistics of the input. The input audio fed to the pitch encoder on the other hand is only perturbed using the latter two processes, whereas the input to the speaker encoder is unaltered. The process of information perturbation ensures that content and pitch features no longer provide speaker-related information, making the input to the speaker encoder unique. This way of controlling the information flow is reported to be significantly useful as it does not suffer from the mismatch problem. It therefore performs well on both CER and SSIM metrics unlike many other information bottleneck approaches \cite{choi_neural_2021}.

\subsubsection{Voice Control and Interpretability} \label{sec:control}
An inherent constraint in the greater part of the analyzed voice conversion models lies in their capacity to only synthesize speech that is either present in the datasets used or defined by a speaker-specific embedding. Manipulating voice in order to create new voice identities or edit specific voice attributes similar to voice transformations, thus remains a challenge for the majority of the present conversion systems. This constraint is often restricted by the low-level representations that most models compose in their information bottleneck \cite{choi2022nansy}. Desirable control features such as pitch, energy or timbre are here entangled either in the latent spaces or in the traditional speaker-dependent and independent embeddings and though some models are further conditioned, this choice does not necessarily force the decoder to learn the mapping of different pitch and energy representations. It is trivial to conclude that the more one decomposes a signal into high-level/interpretable representations, the more one can gain access to the controllability and thereby parameterise the model. Comparatively little research has been devoted to creating speech that sounds like truly novel speakers.

As earlier mentioned only 6 of the works analysed explicitly concentrates on voice control. The work in \cite{choi_neural_2021} is specifically focusing on this. Using an analysis-synthesis procedure they explicitly condition their generation process on separated features counting; linguistics, speaker embedding, pitch and energy which are perturbed using different DSP techniques. Inspired by the source-filter theory, they contrary to other works split the generation process into two; a source and a filter decoder, generating harmonic content and spectral envelopes respectively. They simply incorporate inductive bias in the model by conditioning each generator on features important for the given generation task. This provides both interpretability and formant preserving pitch-shifting capabilities \cite{choi_neural_2021}. The source and filter representations are thereafter summed to represent a mel-spectrogram that is fed to the vocoder of choice. In \cite{xie_end--end_2022} they adopt similar perturbation techniques. However, rather than separating the generation process, they feed the feature embeddings directly to one unified waveform generator. Since the generator synthesises speech directly in the time domain, it can easily control different speech attributes. Lastly, \cite{wang_controllable_2022} introduces a method for controllable speech representation learning based on disentanglement only. This work adheres to a conventional CAE and analysis-synthesis structure, wherein feature encoders and latent embeddings are sequentially arranged prior to their input into the decoder. Nonetheless, through the incorporation of training guidance using reconstruction, content, AIC, and adversarial losses, the authors assert that they achieve a level of disentanglement sufficient for controllability. 

Although the aforementioned works use different structures to achieve interpretability and controllability, they share similarities. Firstly, they in their own way aim to accomplish disentanglement by controlling the information flow. Secondly, they incorporate discriminators for tuning the output quality and naturalness. Lastly, they all incorporate either perturbation or intermediate features obtained by pretrained wav2vec networks for the linguistic and speaker representations. All works offer convincing results, and it is evident that these studies should be considered inspirational sources when designing systems that enable explicit voice control in the future.

\subsubsection{Real-time Constraints}
The efficiency of a voice conversion system during inference is bounded by: the speed of the generator and vocoder used, the speed of converting the utterance between time and frequency domains, and the speed of the encoders specifically the content and speaker encoders. As outlined in this study, most pipelines incorporate encoders based on deep convolutional blocks, recurrent neural networks or large self-supervised models, while rather complex neural vocoders are included to synthesise the acoustic representations. While this is done to ensure the best output quality, it limits the real-time possibilities of the models. It was therefore also found that only 4 works of the studies analysed mention the importance of real-time, stream-able voice conversion \cite{yang_streamable_2022, tanaka_distilling_2023, baas_stargan-zsvc_2020, himawan_jointly_2022}. 

Enabling live one-shot VC is specifically in focus in \cite{yang_streamable_2022}, where each sub-block is carefully designed to allow for streaming. This is among other things done by implementing all convolution and self-attention layers as causal, while all recurrent structures are implemented in a uni-directional manner. Lastly, they adopt a cached sliding-window procedure that processes utterances chunk by chunk making the pipeline applicable for buffer-based computation. The proposed model achieves a real-time factor of 0.37 on a single CPU, and though the work compromises on network structure, results suggest that the model achieves comparable one-shot VC performance with offline solutions \cite{yang_streamable_2022}. Similar approaches are taken in \cite{tanaka_distilling_2023}, where networks also are implemented using causal layers. However, the authors here report degradation due to the "use of causal layers which masks future input information" \cite{tanaka_distilling_2023}. To take account of this, they propose knowledge distillation in which a 'student' encoder, implemented as a streamable structure, learns in conjunction with the more complex, non-causal and non-streamable 'teacher' decoder. However, the work in \cite{tanaka_distilling_2023} does not report any real-time factor.

Broadening our perspective, we discover that the approaches taken in the above studies is followed in similar work that was not indexed by the included databases. In order to adhere to streamable environments, the studies in \cite{ning2023dualvc} and \cite{ning2023dualvc2} also utilise unidirectional recurrent networks, causal convolution layers and knowledge distillation in terms of a teacher-student learning approach. Another interesting approach is the ability to train traditional non-causal networks in order to perform a post-training causal reconfiguration of the trained model as presented in \cite{caillon2022streamable}. Specifically this method provides an interesting ground for real-time neural audio synthesis and voice conversion in the future. 

\section{Limitations}
Although a comprehensive search, guided by a carefully designed keyword list, has been conducted on two of the most relevant databases, most of the review process has been carried out individually by reviewer A and there has been no forward-citation search on the included studies. Additionally, no assessment of the reference list's quality has been carried out, as it was determined that the two included databases upheld a specific scientific standard.

\section{Conclusions}
There is no doubt that voice conversion is a rapidly evolving research area, with numerous challenges and several approaches aimed at addressing them. This paper presented a scoping review of 123 papers in the field of voice conversion. The papers were evaluated using a codebook of 14 codes covering research direction, contributions, methods and deep learning structures employed. Furthermore, we provided an overview of the most commonly used datasets, sampling rates and loss functions. We did this to explore the modern deep learning-powered voice conversion pipeline and chart potential solutions to any encountered obstacles. Currently, the voice conversion community concentrates on addressing the information bottleneck for disentangled speech representation learning. Performance and output quality are often degraded by entanglement in the intermediate feature representations, such as leakage between the content, speaker and prosody embeddings. In this regard, generative structures are utilised to segment the features and synthesise an acoustic representation from the manipulated latent space, while tailored loss functions are employed to fine-tune and regularize the information needed in the given representation. Through these methods, both mismatch problems and information leakage is addressed. Although the proposed techniques are reported to improve the output quality, the actual interpretability of the encoded attributes remains a challenge. In general, the literature presents a minimal focus on generalizability and voice control, reducing voice conversion to an offline identity conversion problem. This constraint may in the future be overcome by specifically controlling the information flow of the system's bottleneck and real-time efficient model structures.

\section{Ethical Statement}
The research activity of this review was considered by the Aalborg University Research Ethics Committee to be ethically justifiable. Nonetheless, it is important to notice that the research covered may provide a ground for technology that can be misused and potentially harm other people. A voice conversion system, both in its offline, controllable or streamable states, may in the wrong hands be used for unethical and disputable applications such as deepfakes or deceptive phone calls. We, therefore, believe that potential systems should be released only to identified and authorized users and that priorities should be put into the development of anti-spoofing technology for identifying fake voices.

\section{Funding}
This work was supported by the 'Industrial PhD grant' from the 'Innovation Fund of Denmark' as well as the virtual reality and augmented reality production studio KhoraVR and its spin-off Heka VR.

\newpage
\bibliographystyle{apalike}
\bibliography{references}  







\section*{Supplementary material}

# Appendix A

| Title | Year | System | Paradigm | Vocoder Used | Adv. Training | Loss function used | Reconstruction Domain | Discriminator Domain | Dataset | Optimizer | Objective Measure | Subjective Measure | Features/Encoder Input | Feature Space | Content Encoder | Timbre Encoder | Type | Goal | Sample rate |
|---|---|---|---|---|---|---|---|---|---|---|---|---|---|---|---|---|---|---|---|
| English Emotional Voice Conversion Using StarGAN Model | 2023 | StarGAN | Unsupervised | Parallel WaveGAN | y | Root mean square error (RMSE) | Acoustic | Hybrid | The dataset used consisted of 720 audio files from 6 speakers | - | MCD, RMSE-Spectrum, RMSE-Prosody | - | Content, style, pitch | General | - | - | Emotion conversion | Improve output | - |
| Non-Parallel Voice Conversion Using Cycle-Consistent Adversarial Networks with Self-Supervised Representations | 2023 | AE | Self-supervised | - | y | Adversarial loss, cycle consistency loss, identity mapping loss | Acoustic | Acoustic | VCC 2016 | - | MCD | - | Content, style | General | Wav2Vec | - | Voice conversion | Disentangle content and speaker | - |
| A Diffeomorphic Flow-Based Variational Framework for Multi-Speaker Emotion Conversion | 2023 | CycleGAN | Unsupervised | WaveNet | y | Adversarial loss, cycle consistency loss | - | Acoustic | VESUS | Adam | RMSE-F0, RMSE-Energy | MOS | Content, emotion | General | BiLSTM | - | Emotion conversion | Disentangle content and prosody | - |
| Voice Conversion Using Learnable Similarity-Guided Masked Autoencoder | 2023 | AE | Self-supervised | - | n | L1 reconstruction loss | Acoustic | - | VCTK | AdamW | SAC, WER | MOS | Content, style | General | CNN | LSGM | Voice conversion | Disentangle content and speaker | - |
| Disentangled Speech Representation Learning for One-Shot Cross-Lingual Voice Conversion Using β-VAE | 2023 | VAE | Unsupervised | Hifi GAN | y | MSE and MAE Reconstruction loss, content KL loss, speaker KL loss | Acoustic | - | VCTK, AISHELL-3 | Adam | EER | MOS | Content, speaker | General | CNN | MLP + ResCNN | Voice conversion | Disentangle content and speaker | - |
| Distilling Sequence-to-Sequence Voice Conversion Models for Streaming Conversion Applications | 2023 | Seq2seq | Self-supervised | Hifi GAN | y | Reconstruction loss, student-teacher loss | Acoustic | - | - | Adam | MCD, RMSE-F0, CER | MOS | Content, speaker | General | CNN | LSTM | Voice conversion | Streamable VC | - |
| Decoupling Speaker-Independent Emotions for Voice Conversion via Source-Filter Networks | 2023 | A-S | Unsupervised | WaveNet | n | Reconstruction loss | Acoustic | - | VCTK | - | MCD, RMSE-F0 | MOS | Content, speaker, pitch, rhythm | General | DNN | DNN | Emotion conversion | - | - |
| Mel-S3R: Combining Mel-spectrogram and self-supervised speech representation with VQ-VAE for any-to-any voice conversion | 2023 | VAE | Self-supervised | MelGAN | n | L1 reconstruction loss, VQ loss | Acoustic | - | LibriTTS, VCTK | - | MCD, RMSE-F0 | MOS | Content, speaker | General | CNN | GST (Global Style Tokens) | Voice conversion | Disentangle content and speech | 16 kHz |
| StyleVC: Non-Parallel Voice Conversion with Adversarial Style Generalization | 2022 | CAE | Unsupervised | HiFi GAN | y | Reconstruction loss, adversarial loss, feature loss | Freq | Freq | VCTK | AdamW | MCD, RMSE-F0, EER, WER | MOS | Content, style | General | Wav2Vec | Meta-StyleSpeech based | Style conversion | Disentangle content and speaker | 16 kHz |
| DISENTANGLING CONTENT AND FINE-GRAINED PROSODY INFORMATION VIA HYBRID ASR BOTTLENECK FEATURES FOR VOICE CONVERSION | 2022 | CAE | Unsupervised | Hifi GAN | y | Reconstruction loss, CE loss, CTC loss | Acoustic | Acoustic | M2VoC | - | MCD, RMSE-F0, RMSE-Energy | ABX, MOS | Content, prosody | General | Conformer | DNN | Voice conversion | Disentangle content and speaker | 16 kHz |
| End-to-End Voice Conversion with Information Perturbation | 2022 | A-S | Unsupervised | Hifi GAN | y | Adversarial loss, feature matching loss, multi-resolution STFT loss | Freq | Hybrid | VCTK | Adam | WER | MOS | Content, speaker, pitch, energy | Parameter | CNN | Res MLP | Voice control | Disentangle content and speaker | 24 kHz |
| Speech Representation Disentanglement with Adversarial Mutual Information Learning for One-shot Voice Conversion | 2022 | A-S | Unsupervised | WaveNet | y | Reconstruction loss, feature loss, adversarial loss | Acoustic | - | VCTK | Adam | MCD, CER, WER, Log-F0 PCC | MOS | Content, speaker, pitch, rhythm | Parameter | CNN + BiLSTM | DNN | Voice conversion | Disentangle content, speaker, pitch, rhythm | 16 kHz |
| FlowCPCVC: A Contrastive Predictive Coding Supervised Flow Framework for Any-to-Any Voice Conversion | 2022 | VAE | Unsupervised | Hifi Gan | y | Reconstruction loss, KL loss, feature matching loss, adversarial loss, CPC loss | Freq | Time | VCTK | Adam | CER, WER, PCC | MOS | Content, speaker | General | CPC | ResNet | Voice conversion | Improve output | 22,05 kHz |
| DRVC: A FRAMEWORK OF ANY-TO-ANY VOICE CONVERSION WITH SELF-SUPERVISED LEARNING | 2022 | AE | Self-supervised | MelGAN | y | Cycle reconstruction loss, adversarial loss, feature matching loss | - | Time | VCC 2018 | Adam | MCD | MOS | Content, style | General | CNN + LSTM | AdaIN-VC | Voice conversion | Disentangle content, speaker | 22,05 kHz |
| CONTROLLABLE SPEECH REPRESENTATION LEARNING VIA VOICE CONVERSION AND AIC LOSS | 2022 | A-S | Unsupervised | Hifi Gan | y | Adversarial loss, reconstruction loss, feature matching loss, AIC loss, content loss | - | Time | VCTK | Adam | L1-Pitch | MOS | Content, speaker, pitch, energy | Parameter | CNN | Resemblyzer | Voice control | Disentangle content, speaker, pitch, energy | 22,05 kHz |
| Timbre Transfer with Variational Auto Encoding and Cycle-Consistent Adversarial Networks | 2022 | VAE | Unsupervised | WaveNet | y | Adversarial loss, cycle consistency loss, KL loss | - | Freq | Flickr 8k Audio, URMP | Adam | SSIM, FAD | - | Content | General | CNN | - | Voice conversion | - | 16 kHz |
| NVC-NET: END-TO-END ADVERSARIAL VOICE CONVERSION | 2022 | C-VAE | Unsupervised | Generator | y | Reconstruction loss, adversarial loss, KL loss, feature matching loss, content preservation loss | Acoustic | Time | VCTK | Adam | - | MOS | Content, speaker | General | CNN | ResCNN | Voice conversion | Disentangle content and speaker | 22,05 kHz |
| TRAINING ROBUST ZERO-SHOT VOICE CONVERSION MODELS WITH SELF-SUPERVISED FEATURES | 2022 | CAE | Self-supervised | Parallel WaveGAN | y | Reconstruction loss, cycle consistency loss, difference loss, speaker id loss | Acoustic | - | LibriTTS, VCTK | - | SVA, EER, CER, MosNet | MOS | Content, speaker | General | Wav2Vec | CNN | Voice conversion | Disentangle content and speaker | - |
| ASGAN-VC: One-Shot Voice Conversion with Additional Style Embedding and Generative Adversarial Networks | 2022 | CAE | Unsupervised | MelGAN | y | Reconstruction loss, content loss, cosine similarity loss | Acoustic | Freq | VCTK | Adam | EER, AUC | MOS | Content, speaker, style | General | CNN + BiLSTM | LSTMDV | Voice conversion | Improve output | 16 kHz |
| High Quality and Similarity One-Shot Voice Conversion Using End-to-End Model | 2022 | CAE | Unsupervised | Hifi GAN | y | Reconstruction loss, feature loss, adversarial loss, feature matching loss | Time | Time | VCTK | AdamW | WER | XAB, MOS | Content, speaker | General | Other | ECAPA-TDNN | Voice conversion | Disentangle content and speaker | 16 kHz |
| Non-Parallel Voice Conversion Based on Free-Energy Minimization of Speaker-Conditional Restricted Boltzmann Machine | 2022 | RBM | Unsupervised | WORLD | y | Adversarial loss, cycle consistency loss, identity mapping loss, and classification loss | - | - | CMU-ARCTIC | - | MCD | MOS | Content, speaker | General | - | - | Voice conversion | - | 16 kHz |
| Pyramid Attention CycleGAN for Non-Parallel Voice Conversion | 2022 | CycleGAN | Self-supervised | MelGAN | y | Adversarial loss, identity mapping loss | Acoustic | Acoustic | VCC 2018 | Adam | MCD | XAB, MOS | Content | General | Other | - | Voice conversion | Improve content embedding | 22,05 kHz |
| IMPROVING RECOGNITION-SYNTHESIS BASED ANY-TO-ONE VOICE CONVERSION WITH CYCLIC TRAINING | 2022 | Cyclic CAE | Unsupervised | Parallel WaveGAN | y | Reconstruction loss, adversarial feature loss, cyclic loss | Acoustic | - | Other | Adam | MCD, RMSE-F0 | MOS | Content, speaker | General | ASR | BiLSTM | Voice conversion | Improve content embedding | - |
| C-CycleTransGAN: A Non-parallel Controllable Cross-gender Voice Conversion Model with CycleGAN and Transformer | 2022 | Cyclic CAE | Self-supervised | WORLD | y | Consistency loss, identity loss, and adversarial loss | Time | Acoustic | VCC 2020 | Adam | MCD | MOS | Content, speaker | Parameter | CNN | - | Voice control | Improve speaker embedding | - |
| Voice Frequency Synthesis using VAWGAN based Amplitude Scaling for Emotion Transformation | 2022 | VAE | Unsupervised | WORLD | y | - | - | Acoustic | RAVDESS | - | MCD, LSD | - | Content, pitch | Parameter | - | - | Emotion conversion | Disentangle prosody and content | 48 kHz |
| Duration Controllable Voice Conversion via Phoneme-Based Information Bottleneck | 2022 | CAE | Unsupervised | WaveNet | y | Reconstruction loss, adversarial loss, phoneme loss, duration loss | Acoustic | - | VCTK | Adam | MCD, RMSE-F0 | MOS | Content, speaker | General | Other | LSTM | Rhythm conversion | Disentangle content, speaker and style | 22,05 kHz |
| U-GAT-VC: UNSUPERVISED GENERATIVE ATTENTIONAL NETWORKS FOR NON-PARALLEL VOICE CONVERSION | 2022 | AE | Unsupervised | - | y | Adversarial loss, identity mapping loss, cycle consistency loss | Acoustic | Acoustic | VCC 2018 | - | FDSD, KDSD | ABX | Content | General | CNN | - | Voice conversion | Improve output | 22,05 kHz |

| Title | Year | Model | Training | Vocoder | (col) | Loss | Feature | (col2) | Dataset | Optimizer | Metric | Subjective | Features | Conversion | Content Enc | Speaker Enc | Task | Contribution | Sample Rate |
|---|---|---|---|---|---|---|---|---|---|---|---|---|---|---|---|---|---|---|---|
| Zero-shot Voice Conversion via Self-supervised Prosody Representation Learning | 2022 | AE | Self-supervised | PWGAN | n | Reconstruction loss, CPC loss, prosody loss | Acoustic | - | VCTK | Adam | Speaker Verification | MOS | Content, speaker, pitch, energy | General | CPC | DNN | Voice conversion | Disentangle content and speaker | 22,05 kHz |
| Non-Parallel Voice Conversion System Using An Auto-Regressive Model | 2022 | AE | Unsupervised | LPCNet | n | Mean Square Error (MSE) | Time | - | CMU-ARCTIC | Adam | MCD | MOS | Content | General | ASR | - | Voice conversion | Improve output | 16 kHz |
| Disentanglement of Emotional Style and Speaker Identity for Expressive Voice Conversion | 2022 | A-S | Unsupervised | Parallel WaveGAN | n | Reconstruction loss, feature loss | Acoustic | - | ESD | Adam | MCD | ABX, MOS | Content, style, speaker | Parameter | DNN | DNN | Voice conversion | Disentangle content, rhythm and speaker | 16 kHz |
| SPEECHSPLIT2.0: UNSUPERVISED SPEECH DISENTANGLEMENT FOR VOICE CONVERSION WITHOUT TUNING AUTOENCODER BOTTLENECKS | 2022 | CAE | Unsupervised | - | n | Reconstruction loss | Acoustic | - | VCTK | Adam | CER | MOS | Content, rhythm, pitch | Parameter | CNN | One-hot vector | Voice conversion | Disentangle content, speaker, pitch | - |
| A COMPARISON OF DISCRETE AND SOFT SPEECH UNITS FOR IMPROVED VOICE CONVERSION | 2022 | AE | Self-supervised | WaveNet, Hifi GAN | n | Contrastive loss | - | - | LJSpeech | - | WER, PER | MOS | Content | General | CPC | - | Voice conversion | Improve content embedding | 16 kHz |
| SIG-VC: A SPEAKER INFORMATION GUIDED ZERO-SHOT VOICE CONVERSION SYSTEM FOR BOTH HUMAN BEINGS AND MACHINES | 2022 | CAE | Unsupervised | MelGAN | n | Reconstruction loss, post-net reconstruction loss, std loss, speaker recognition loss | Acoustic | - | VCTK | Adam | Cos Sim | MOS | Content, speaker | General | TDNN | ECAPA-TDNN | Voice conversion | Disentangle content and speaker | 16 kHz |
| One-shot Voice Conversion For Style Transfer Based On Speaker Adaptation | 2022 | CAE | Unsupervised | LPCNet | n | Reconstruction loss, Speaker cross-entropy loss | Acoustic | - | VCTK, CMU-ARCTIC | Adam | PCC-Features | MOS | Content, speaker, prosody | General | Other | TDNN-F | Voice conversion | Improve output | 16 kHz |
| A Study on Low-Latency Recognition-Synthesis-Based Any-to-One Voice Conversion | 2022 | CAE | Unsupervised | LPCNet | n | MSE reconstruction loss, MMI loss, speaker id loss | Acoustic | - | iFlytek | - | WER | MOS | Content, speaker | General | ASR | BN + LSTM | Voice conversion | Low-latency | - |
| ASSEM-VC: REALISTIC VOICE CONVERSION BY ASSEMBLING MODERN SPEECH SYNTHESIS TECHNIQUES | 2022 | CAE | Unsupervised | Hifi GAN | n | - | - | - | LibriTTS, VCTK | - | - | MOS, DMOS | Content, speaker, pitch | Parameter | ASR | - | Voice conversion | Disentangle content and speaker | - |
| Towards Improved Zero-shot Voice Conversion with Conditional DSVAE | 2022 | C-VAE | Unsupervised | Hifi GAN | n | NLL Reconstruction loss, KL feature loss | Acoustic | - | VCTK | Adam | ACC-Phoneme | MOS | Content, speaker | General | BiLSTM | BiLSTM | Voice conversion | Improve content embedding | - |
| A New Spoken Language Teaching Tech: Combining Multi-attention and AdaIN for One-shot Cross Language Voice Conversion | 2022 | VAE | Self-supervised | Hifi GAN | n | Reconstruction loss, KL loss | Acoustic | - | AISHELL-3, VCTK | Adam | MosNet, TA, PER, PDR | MOS | Content, speaker | General | CNN | CNN | Voice conversion | Improve content embedding | - |
| Streamable Speech Representation Disentanglement and Multi-Level Prosody Modeling for Live One-Shot Voice Conversion | 2022 | A-S | Unsupervised | WaveRNN | n | Reconstruction loss, VQ loss, CPC loss, MI loss | Acoustic | - | Librispeech, VCTK | Adam | CER, WER | MOS | Content, speaker, pitch | Parameter | CPC | UPLF | Voice conversion | Improve content embedding | 16 kHz |
| ROBUST DISENTANGLED VARIATIONAL SPEECH REPRESENTATION LEARNING FOR ZERO-SHOT VOICE CONVERSION | 2022 | Seq2seq | Self-supervised | Griffin-Lim, WaveNet | n | Reconstruction loss, KL loss | Acoustic | - | TIMIT, VCTK | Adam | EER | MOS | Content, speaker | General | BiLSTM + RNN | BiLSTM + RNN | Voice conversion | Disentangle content and speaker | - |
| Jointly Trained Conversion Model With LPCNet for Any-to-One Voice Conversion Using Speaker-Independent Linguistic Features | 2022 | CAE | Unsupervised | LPCNet | n | MSE Reconstrcution loss, LPCNet loss | Acoustic | - | LibriVox | Adam | Cos Sim | ABX, MOS | Content, speaker | General | ASR | DNN | Voice conversion | Disentangle content and speaker | 16 kHz |
| Parallel voice conversion with limited training data using stochastic variational deep kernel learning | 2022 | AE | Unsupervised | WORLD | n | MSE Reconstruction loss | Time | - | VCTK | Adam | MCD | MOS | Spectral envelope, pitch | Parameter | Other | - | Voice conversion | Improve output | 16 kHz |
| Accent Conversion using Pre-trained Model and Synthesized Data from Voice Conversion | 2022 | Seq2seq | Self-supervised | WaveGlow | n | Reconstruction loss, CTC loss | Acoustic | - | CMU-ARCTIC, L2-ARCTIC | Adam | WER, ACC | MOS | Content, accent | General | BiLSTM | - | Accent conversion | Disentangle content and prosody | 16 kHz |
| Accentron: Foreign accent conversion to arbitrary non-native speakers using zero-shot learning | 2022 | Seq2seq | Unsupervised | WaveRNN | n | Reconstruction loss, CE loss | Acoustic | - | Librispeech, Speech Accent Archive, VoxCeleb1 | Adam | - | MOS | Content, speaker, style (accent) | General | TDNN | CNN | Accent conversion | Disentangle content and accent | 16 kHz |
| Foreign Accent Conversion using Concentrated Attention | 2022 | AE | Unsupervised | WaveGlow | n | Reconstruction loss, CTC loss, CE loss | Acoustic | - | L2-ARTIC | - | MSD | MOS | Content | General | ASR | - | Accent conversion | Improve output | 16 kHz |
| Enhancing Zero-Shot Many to Many Voice Conversion via Self-Attention VAE | 2022 | VAE | Self-supervised | WaveNet | n | Reconstruction loss, Negative Evidence Lower Bound (ELBO), KL loss | Acoustic | - | VCTK | - | MOSNet, SPC-ACC | - | Content, speaker | General | CNN + BiLSTM | - | Voice conversion | Improve content embedding | - |
| Enhancing Zero-Shot Many to Many Voice Conversion via Self-Attention VAE with Structurally Regularized Layers | 2022 | VAE | Unsupervised | WaveNet | n | ELBO | Acoustic | - | VCTK | - | MOSNet, SPC-ACC | - | Content | General | CNN + LSTM | - | Voice conversion | Improve output | - |
| Adversarially Learning Disentangled Speech Representations for Robust Multi-factor Voice Conversion | 2021 | A-S | Unsupervised | MelGAN | y | Adversarial loss, reconstruction loss | Acoustic | - | VCTK | Adam | MCD | MOS | Content, speaker, pitch, rhythm | Parameter | DNN | DNN | Voice conversion | Disentangle, content, speaker, pitch, rhythm | 16 kHz |
| Two-Pathway Style Embedding for Arbitrary Voice Conversion | 2021 | CAE | Unsupervised | MelGAN | y | Reconstruction loss, content loss | - | Time | VCTK | Adam | Cos Sim | MOS | Content, style | General | CNN | CNN | Voice conversion | Disentangle content and speaker | 22,05 kHz |
| Towards Unseen Speakers Zero-Shot Voice Conversion with Generative Adversarial Networks | 2021 | GAN | Unsupervised | WaveNet | y | Reconstruction loss, adversarial loss, feature matching loss, KL loss | Acoustic | Freq | VCTK | SGD, Adam | MCD | MOS | Content, timbre | General | CNN + BiLSTM | LSTM | Voice conversion | Disentangle content and speaker | 16 kHz |
| An Improved StarGAN for Emotional Voice Conversion: Enhancing Voice Quality and Data Augmentation | 2021 | CAE | Unsupervised | WORLD | y | Reconstruction loss, feature loss | Acoustic | Freq | IEMOCAP | Adam | - | MOS | Content, emotion | General | - | - | Emotion conversion | Improve emotion embedding | 16 kHz |
| VAW-GAN FOR DISENTANGLEMENT AND RECOMPOSITION OF EMOTIONAL ELEMENTS IN SPEECH | 2021 | VAE | Unsupervised | WORLD | y | Reconstruction loss, KL loss, adversarial loss | Acoustic | Acoustic | - | RMSProp | RMSE, PCC | ABX, MOS | Content, emotion, pitch | Parameter | CNN | - | Emotion conversion | Disentangle content, prosody and emotion | 16 kHz |
| SEEN AND UNSEEN EMOTIONAL STYLE TRANSFER FOR VOICE CONVERSION WITH A NEW EMOTIONAL SPEECH DATASET | 2021 | C-VAE | Unsupervised | WORLD | y | - | - | Acoustic | ESD | - | MCD | ABX, MOS | Content, emotion, pitch | Parameter | CNN | - | Emotion conversion | Disentangle content and emotion | 16 kHz |
| Accent and Speaker Disentanglement in Many-to-many Voice Conversion | 2021 | CAE | Unsupervised | melLPCNet | y | Reconstruction loss, adversarial loss | Acoustic | - | - | Adam | - | MOS | Content, speaker | General | ASR | - | Accent conversion | Propose to use adversarial training to better disentangle the speaker and accent information | 16 kHz |
| WINVC: One-Shot Voice Conversion with Weight Adaptive Instance Normalization | 2021 | StarGAN | Unsupervised | Parallel WaveGAN | y | Adversarial loss, cycle consistency loss, identity loss, speaker embedding loss | Acoustic | Acoustic | VCTK | - | - | ABX, MOS, sMOS | Content, speaker | General | CNN + IN | CNN | Voice conversion | Disentangle content and speaker | - |

| Title | Year | Model | Learning | Vocoder | Parallel | Loss | Input | Output | Dataset | Optimizer | Objective Metrics | Subjective Metrics | Disentangled features | Speaker embedding | Encoder | Decoder | Task | Contribution | Sample rate |
|---|---|---|---|---|---|---|---|---|---|---|---|---|---|---|---|---|---|---|---|
| Cross-Lingual Voice Conversion With Controllable Speaker Individuality Using Variational Autoencoder and Star Generative Adversarial Network | 2021 | C-VAE | Unsupervised | Parallel WaveGAN | y | Adversarial loss, KL loss, Classification loss, cycle consistency loss, speaker id loss | Acoustic | Acoustic | VCTK, JVS | Adam | Mean-F0, MS | ABX, MOS | Content, speaker, pitch | Parameter | CNN | Codebook | Voice conversion | Disentangle content and speaker | 24 kHz |
| Effects of Sinusoidal Model on Non-Parallel Voice Conversion with Adversarial Learning | 2021 | A-S | Unsupervised | Source Filter | y | Reconstruction loss, Adversarial loss, Classification loss | Acoustic | Acoustic | VCTK | Adam | RMS-LSD, PDM | SSIM | Pitch, MVG, Spectral envelope | Parameter | Other | - | Voice conversion | Improve output quality | 16 kHz |
| Neural Analysis and Synthesis: Reconstructing Speech from Self-Supervised Representations | 2021 | A-S | Self-supervised | HiFi GAN | y | Speaker conditioned adversarial loss, reconstruction loss | Acoustic | Freq | VCTK, LibriTTS | Adam | CER | MOS, DMOS | Content, speaker, pitch, energy | Parameter | Wav2Vec | Wav2Vec | Voice control | Disentangle content, speaker, pitch, energy for explicit control | 22,05 kHz |
| StarGAN-VC+ASR: StarGAN-based Non-Parallel Voice Conversion Regularized by Automatic Speech Recognition | 2021 | StarGAN | Unsupervised | - | y | adversarial, cycle consistency loss, identity mapping loss, domain classification loss | - | Acoustic | ATR | Adam | CER | MOS | Content | General | ASR | - | Voice conversion | Improve content embedding | 20 kHz |
| Many-to-Many Unsupervised Speech Conversion from Nonparallel Corpora | 2021 | C-VAE | Unsupervised | - | y | Reconstruction loss, cycle reconstruction loss, cycle feature loss, feature reconstruction loss | Acoustic | Acoustic | VCC 2016 | Adam | MCD | ABX, MOS | Content, speaker | General | CNN + GRU | CNN | Voice conversion | Improve speaker embedding | 16 kHz |
| S2VC: A Framework for Any-to-Any Voice Conversion with Self-Supervised Pretrained Representations | 2021 | CAE | Self-supervised | Universal Neural Vocoder (PWNet) | n | Log Mel spectrogram L1 loss | Acoustic | - | VCTK | AdamW | MOSNet | MOS | Content, speaker | General | MLP | CNN | Voice conversion | Disentangle content and speaker | 16 kHz |
| ONE-SHOT VOICE CONVERSION BASED ON SPEAKER AWARE MODULE | 2021 | CAE | Unsupervised | LPCNet | n | L1 Reconstruction loss | Acoustic | - | AISHELL-1 | Adam | MCD | MOS | Content, speaker | General | DNN | Speaker Aware Module (SAM) | Voice conversion | Disentangle content and speaker | - |
| VQMIVC: Vector Quantization and Mutual Information-Based Unsupervised Speech Representation Disentanglement for One-shot Voice Conversion | 2021 | A-S | Unsupervised | Parallel WaveGAN | n | Reconstruction loss, VQCPC loss, and MI loss | Acoustic | - | VCTK | Adam | CER, WER | MOS | Content, speaker, pitch | Parameter | DNN | CNN | Voice conversion | Disentangle content and speaker | 16 kHz |
| ZERO-SHOT VOICE CONVERSION WITH ADJUSTED SPEAKER EMBEDDINGS AND SIMPLE ACOUSTIC FEATURES | 2021 | A-S | Unsupervised | WaveNet | n | Reconstruction loss, feature loss | Acoustic | - | VCTK | Adam | Fake Speech Detection | MOS | Content, speaker, pitch | Parameter | CNN + BiLSTM | CNN + LSTM | Voice conversion | Improve speaker embedding | 16 kHz |
| END-TO-END ZERO-SHOT VOICE CONVERSION USING A DDSP VOCODER | 2021 | CAE | Unsupervised | DDSP | n | Reconstruction loss | Freq | - | VCTK | Adam | MSE, MSL | MOS | Content, speaker, pitch | Parameter | BiLSTM | LSTM | Voice conversion | Disentangle content and speaker | 16 kHz |
| ANY-TO-ONE SEQUENCE-TO-SEQUENCE VOICE CONVERSION USING SELF-SUPERVISED DISCRETE SPEECH REPRESENTATIONS | 2021 | Seq2seq | Self-supervised | Parallel WaveGAN | n | Contrastive loss | - | - | M-AILABS | - | MCD, WER, CER | MOS | Content | General | Wav2Vec | - | Voice conversion | Improve output | 16 kHz |
| Many-to-Many Voice Conversion based Feature Disentanglement using Variational Autoencoder | 2021 | VAE | Unsupervised | WaveNet | n | Reconstruction loss, Evidence Lower Bound (ELBO) | Acoustic | - | VCTK | Adam | MCD | MOS | Content, speaker | General | CNN + BiLSTM | CNN + BiLSTM | Voice conversion | Disentangle content and speaker | 16 kHz |
| Many-to-Many Unsupervised Speech Conversion from Nonparallel Corpora | 2021 | CAE | Unsupervised | HiFi GAN | n | MAE, CE loss | Acoustic | - | VCTK, Librispeech | - | WER, Cosine Distance | MOS | Content, speaker | General | TDNN | Speaker Statistics | Voice conversion | Improve speaker embedding | - |
| Connectionist temporal classification loss for vector quantized variational autoencoder in zero-shot voice conversion | 2021 | VQ-VAE | Unsupervised | MelGAN | n | MSE reconstruction loss, CTC loss, VQ loss | Acoustic | - | VCTK, CMU-ARCTIC | - | Q-Perplexity, SPC-ACC, VQ Continuity Score | MOS | Content, speaker | General | CNN + BiGRU | Global Style Tokens (GST) | Voice conversion | Improve content embedding | 16 kHz |
| Sequence-to-Sequence Emotional Voice Conversion With Strength Control | 2021 | Seq2seq | Unsupervised | Parallel WaveGAN | n | Reconstruction loss, BCE, guided attention loss | Acoustic/Sequences | - | Korean speech corpus | Adam | MCD, VDE, GPE, FFE | MOS | Content, speaker, style | General | CNN + BiLSTM | One-hot vector | Emotion conversion | Disentangle content and style | 22,05 kHz |
| Non-parallel any-to-many voice conversion by replacing speaker statistics | 2021 | CAE | Unsupervised | Hifi GAN | n | MAE | Acoustic | - | Librispeech, VCTK | - | WER, Cosine Distance | MOS | Content, speaker | General | TDNN | Statistics | Voice conversion | Improve output | - |
| Improving the Speaker Identity of Non-Parallel Many-to-Many Voice Conversion with Adversarial Speaker Recognition | 2020 | CAE | Unsupervised | WaveRNN | y | Reconstruction loss, CE loss, adversarial speaker classification loss | Acoustic | - | VCTK | Adam | MCD | VSS, MOS | Content, speaker | General | LSTM | D-Vector | Voice conversion | Improve content embedding | 16 kHz |
| Nonparallel Emotional Speech Conversion Using VAE-GAN | 2020 | VAE | Unsupervised | WORLD | y | Adversarial loss, cycle consistency loss, identity mapping loss, KL loss | Acoustic | Acoustic | IEMOCAP | Adam | - | MOS, Emotion Conversion Ability | Content, style | Parameter | CNN + IN | - | Emotion conversion | Disentangle content and emotion | 16 kHz |
| MANY-TO-MANY VOICE CONVERSION USING CONDITIONAL CYCLE-CONSISTENT ADVERSARIAL NETWORKS | 2020 | C-GAN | Unsupervised | WORLD | y | Adversarial loss, Cycle consistency loss, identity mapping loss | Acoustic | Acoustic | VCC 2018 | Adam | MCD, MSD | MOS | Content, speaker, pitch | Parameter | CNN | - | Voice conversion | Improve output | 22,05 kHz |
| Non-parallel Many-to-many Voice Conversion with PSR-StarGAN | 2020 | StarGAN | Unsupervised | WORLD | y | Adversarial loss, Cycle consistency loss, identity mapping loss, Classification loss | Acoustic | Freq | VCC 2018 | - | - | MOS | Content, speaker | Parameter | CNN | - | Voice conversion | Improve output | - |
| Non-parallel emotion conversion using a deep-generative hybrid network and an adversarial pair discriminator | 2020 | CycleGAN | Unsupervised | CycleGAN | y | Cycle consistency loss, KL loss, spectral loss | Acoustic | - | VESUV | Adam | MAE-F0 | MOS | Content, timbre, pitch | General | ResCNN | - | Emotion conversion | Improve output | - |
| StarGAN-ZSVC: Towards Zero-Shot Voice Conversion in Low-Resource Contexts | 2020 | C-GAN | Unsupervised | WaveGlow | y | Reconstruction loss, aversarial loss, cyclic loss | Acoustic | Acoustic | VCC 2018 | Adam | MAE, MSE | MOS | Cointent, speaker | General | CNN | GRU | Voice conversion | Disentangle content and speaker | 22,05 kHz |
| GAZEV: GAN-based zero-shot voice conversion over non-parallel speech corpus | 2020 | C-GAN | Unsupervised | - | y | Adversarial loss, classification loss, cycle consistency loss, identity mapping loss | Acoustic | Acoustic | VCTK | Adam | - | MOS | Content, speaker | General | ResCNN | MLP | Voice conversion | Improve output | - |
| Unsupervised Representation Disentanglement Using Cross Domain Features and Adversarial Learning in Variational Autoencoder Based Voice Conversion | 2020 | CAE | Unsupervised | WORLD | y | Reconstruction loss, KL loss, adversarial loss, similarity loss, speaker classification | Acoustic | - | VCC 2018 | Adam | MCD, GV, MS | MOS | Content, speaker, pitch | Parameter | CNN | CNN | Voice conversion | Disentangle content and speaker | 22,05 kHz |
| Cyclic spectral modeling for unsupervised unit discovery into voice conversion with excitation and waveform modeling | 2020 | Cyclic VAE | Unsupervised | WaveNet | y | Reconstruction loss, cycle loss, KL loss | Acoustic | - | ZeroSpeech 2020 | Adam | MCD, RMSE, Cos Sim, CER | ABX, MOS | Content, speaker, pitch | Parameter | - | - | Voice conversion | Spectral modelling | 16 kHz |
| Non-Parallel Sequence-to-Sequence Voice Conversion With Disentangled Linguistic and Speaker Representations | 2020 | Seq2seq | Unsupervised | WaveNet | y | Reconstruction loss, adversarial speaker classification, feature loss | Acoustic | - | VCTK, CMU-ARCTIC | Adam | MCD, RMSE-F0, VUV Error, F0 Correlation, DDUR | MOS | Content, speaker | General | BiLSTM | BLSTM | Voice conversion | Improve output | - |

| Title | Year | Model | Training | Vocoder | Parallel | Loss | Feature | Feature2 | Dataset | Optimizer | Metric1 | Metric2 | Conditioning | Type | Network1 | Network2 | Task | Goal | Sample Rate |
|---|---|---|---|---|---|---|---|---|---|---|---|---|---|---|---|---|---|---|---|
| AN IMPROVED FRAME-UNIT-SELECTION BASED VOICE CONVERSION SYSTEM WITHOUT PARALLEL TRAINING DATA | 2020 | AE | Unsupervised | STRAIGHT | n | MSE | Acoustic | - | WeChat, CMU-ARCTIC | ASGD | MSE | MOS | Content, pitch | General | LSTM | - | Voice conversion | Improve output | - |
| Emotional Voice Conversion Using a Hybrid Framework With Speaker-Adaptive DNN and Particle-Swarm-Optimized Neural Network | 2020 | A-S | Unsupervised | STRAIGHT | n | Mean Square Error (MSE) | Time | - | EmoDB, IITKGP, SAVEE | Particle Swarm Optimization (PSO) | MCD, RMSE-F0, PESQ | CMOS | Speaker, pitch, prosody (MGCEP) | Parameter | - | DNN | Emotion conversion | Improve output | 16 kHz |
| Improved Zero-Shot Voice Conversion Using Explicit Conditioning Signals | 2020 | A-S | Unsupervised | WaveRNN | n | Reconstruction loss, latent regressor loss | Acoustic | - | VCTK | Adam | L1-Melspech | MOS | Content, speaker, pitch, energy | Parameter | LSTM | DNN | Voice control | Disentangle content, speaker, pitch, energy for explicit control | 16 kHz |
| ConvS2S-VC: Fully Convolutional Sequence-to-Sequence Voice Conversion | 2020 | Seq2seq | Unsupervised | WORLD | n | Reconstruction loss, diagonal attention loss (DAL), orthogonal attention loss (OAL) | Acoustic | - | CMU-ARCTIC | Adam | MCD, Log-F0 Correlation, LDR | MOS | Content, speaker, pitch | General | - | - | Voice conversion | Convert features | 16 kHz |
| Transferring Source Style in Non-Parallel Voice Conversion | 2020 | Seq2seq | Self-supervised | WaveRNN | n | Reconstruction loss, CTC loss, feature loss | Acoustic | - | - | - | PER | MOS | Content, speaker, style | General | ASR | GST (Global Style Tokens) | Style conversion | Disentangle content and style | 24 kHz |
| Unsupervised Acoustic Unit Representation Learning for Voice Conversion using WaveNet Auto-encoders | 2020 | VQ CAE | Unsupervised | WaveNet | n | VQ loss | - | - | ZeroSpeech 2020 | Adam | CER | ABX, MOS | Content, speaker | General | CNN + IN | CNN | Voice conversion | Improve output | - |
| MULTI-SPEAKER AND MULTI-DOMAIN EMOTIONAL VOICE CONVERSION USING FACTORIZED HIERARCHICAL VARIATIONAL AUTOENCODER | 2020 | VAE | Unsupervised | Griffin-Lim | n | Reconstruciton loss, KL loss, Cycle consistency loss | Acoustic | Acoustic | - | - | - | MOS | Content, emotion | General | LSTM | - | Emotion conversion | Disentangle content and emotion | - |
| ONE-SHOT VOICE CONVERSION BY VECTOR QUANTIZATION | 2020 | VQ AE | Unsupervised | Griffin-Lim | n | Reconstruction loss, latent loss | Time | - | VCTK | Adam | SPC-ACC, L1 | MOS | Content, speaker | General | CNN + VQ | CNN | Voice conversion | Improve output | 24 kHz |
| Multi-speaker emotion conversion via latent variable regularization and a chained encoder-decoder-predictor network | 2020 | CAE | Unsupervised | WORLD | n | Reconstruction loss, feature loss | Acoustic | - | VESUV | Adam | - | MOS | Spectrum, pitch | General | ResCNN | - | Emotion conversion | Improve output | - |
| DNN-Based Cross-Lingual Voice Conversion Using Bottleneck Features | 2020 | CAE | Unsupervised | WORLD | n | MSE | Freq | - | Telugu | RMSProp | - | - | Spectrum, speaker, pitch | Parameter | MLP | - | Voice conversion | Improve output | 16 kHz |
| Multi-Task WaveRNN With an Integrated Architecture for Cross-Lingual Voice Conversion | 2020 | AE | Unsupervised | WORLD, WaveRNN | n | CE loss | Time | - | VCTK, CMU-ARCTIC, VCC 2016, VCC 2018 | - | MCD, RMSE | MOS | Content | Parameter | LSTM + GRU | - | Cross-lingual conversion | Improve output | 16 kHz |
| Cross-Lingual Voice Conversion using a Cyclic Variational Auto-encoder and a WaveNet Vocoder | 2020 | Cyclic VAE | Unsupervised | WaveNet | n | VAE loss, KL loss, feature loss | - | - | VCC 2018 | Adam | MCD | MOS | Content, timbre, pitch | General | RNN | - | Cross-lingual conversion | - | - |
| VQVC plus : One-Shot Voice Conversion by Vector Quantization and U-Net architecture | 2020 | AE | Unsupervised | MelGAN | n | Reconstruction loss, latent loss | Acoustic | - | VCTK | Adam | - | MOS | Content | General | CNN | - | Voice conversion | Disentangle content and speaker | 22,05 kHz |
| Novel Adaptive Generative Adversarial Network for Voice Conversion | 2019 | GAN | Unsupervised | AHOCODER | y | Adversarial loss, cycle consistency loss, style transfer loss, content loss | Acoustic | Acoustic | VCC 2018 | Adam | MCD | MOS | Content, style | General | MLP | - | Style conversion | Disentangle content and style | - |
| Fast Learning for Non-Parallel Many-to-Many Voice Conversion with Residual Star Generative Adversarial Networks | 2019 | GAN | Unsupervised | WORLD | y | Reconstruction loss, adversarial loss, identity mapping loss, classification loss | Acoustic | - | CMU-ARCTIC | - | - | AB, ABX | MCC, Pitch | Parameter | CNN | - | Voice conversion | Improve output | 16 kHz |
| Voice Conversion from Tibetan Amdo Dialect to Tibetan U-tsang Dialect Based on Generative Adversarial Networks | 2019 | GAN | Unsupervised | Generator | y | MSE, Adversarial loss | Acoustic | - | - | - | - | ABX, MOS | MGC, pitch, BAP | Parameter | - | - | Style conversion | Improve output | 16 kHz |
| Cyclegan-VC2: Improved Cyclegan-based Non-parallel Voice Conversion | 2019 | CycleGAN | Unsupervised | WORLD | y | Adversarial loss, cycle consistency loss, identity mapping loss | Acoustic | Acoustic | VCC 2018 | Adam | MCD, MSD | MOS | Spectral envelope, pitch | Parameter | CNN | - | Voice conversion | Improve output | 22,05 kHz |
| StarGAN-VC2: Rethinking Conditional Methods for StarGAN-Based Voice Conversion | 2019 | CycleGAN | Unsupervised | WORLD | y | Adversarial loss, classification loss, cycle consistency loss, identity mapping loss | Acoustic | Acoustic | VCC 2018 | Adam | MCD, MSD | MOS | Content, speaker, pitch | Parameter | CNN | One-hot vector | Voice conversion | Improve output | 22,05 kHz |
| Adversarially Trained Autoencoders for Parallel-data-free Voice Conversion | 2019 | AE | Unsupervised | Griffin-Lim | y | Reconstruction loss, classification loss | Acoustic | - | CSTR | - | - | AB, ABX, MOS | Content | General | CNN | - | Voice conversion | Disentangle content and speaker | - |
| A MODULARIZED NEURAL NETWORK WITH LANGUAGE-SPECIFIC OUTPUT LAYERS FOR CROSS-LINGUAL VOICE CONVERSION | 2019 | A-S | Unsupervised | WOLRD | n | MSE | Acoustic | - | VCC 2016 | - | MCD | AB, MOS | Content, speaker, pitch | Parameter | BiLSTM | i-vector | Cross-lingual conversion | Improve output | 16 kHz |
| Refined WaveNet Vocoder for Variational Autoencoder Based Voice Conversion | 2019 | VAE | Unsupervised | WaveNet | n | VAE loss, KL loss | - | - | VCC 2018 | Adam | MCD | MOS | Spectral envelope, pitch | Parameter | CNN | - | Voice conversion | Improve output | 22,05 kHz |
| Jointly Trained Conversion Model and WaveNet Vocoder for Non-parallel Voice Conversion using Mel-spectrograms and Phonetic Posteriorgrams | 2019 | AE | Unsupervised | WaveNet | n | Reconstruction loss, WaveNet loss | Acoustic | - | CMU-ARCTIC | Adam | MCD, RMSE-F0 | ABX | Pitch, VUV, MFCC | Parameter | ASR | - | Voice conversion | Improve output | 16 kHz |
| One-shot Voice Conversion with Disentangled Representations by Leveraging Phonetic Posteriorgrams | 2019 | CAE | Unsupervised | - | n | Reconstruction loss, feature loss | Time | - | TIMIT | Adam | - | ABX, MOS | Content, speaker | General | ASR | GRU | Voice conversion | Improve speaker embedding | 16 kHz |
| Foreign Accent Conversion by Synthesizing Speech from Phonetic Posteriorgrams | 2019 | AE | Unsupervised | WaveGlow | n | Reconstruction loss, CE loss | Acoustic | - | Librispeech | Adam | - | MOS | Content | General | CNN + BiLSTM | - | Accent conversion | Improve output | 16 kHz |
| Semi-supervised voice conversion with amortized variational inference | 2019 | VAE | Semi-supervised | WORLD | n | VAE loss | - | - | CMU-ARCTIC | - | MCD | MOS | Spectral envelope, pitch | Parameter | LSTM | - | Voice conversion | Improve output | 16 kHz |
| Investigation of F0 conditioning and fully convolutional networks in variational autoencoder based voice conversion | 2019 | VAE | Unsupervised | WORLD | n | Reconstruction loss, cross-domain reconstruction loss, similarity loss, KL loss | Acoustic | - | VCC 2018 | Adam | MCD | MOS | Spectral envelope, mcc, pitch | Parameter | CNN | - | Voice conversion | Improve output | 22,05 kHz |
| Sequence-to-Sequence Acoustic Modeling for Voice Conversion | 2019 | Seq2seq | Unsupervised | WaveNet | n | Reconstruction loss, CE loss | Acoustic | - | CMU-ARCTIC | Adam | MCD, RMSE-F0, DDUR | MOS | Content | General | BiLSTM | - | Voice conversion | Improve output, duration conversion | 16 kHz |
| A BLSTM and WaveNet-Based Voice Conversion Method With Waveform Collapse Suppression by Post-Processing | 2019 | AE | Unsupervised | WaveNet | n | Reconstruction loss, CE loss | - | - | VCC 2018 | Adam | MCD | MOS | Content | Parameter | BiLSTM | - | Voice conversion | Improve output | 22,05 kHz |
| Non-parallel Voice Conversion with Controllable Speaker Individuality using Variational Autoencoder | 2019 | C-VAE | Unsupervised | WORLD | n | VAE loss, modulation spectrum loss | - | - | VCC 2018 | - | - | MOS | Content, spectral envelope, pitch | Parameter | CNN | MLP | Voice control | Disentangle content and speaker | - |

| Title | Year | Model | Training | Vocoder | GAN | Loss | Features domain | Distribution alignment | Dataset | Optimizer | Objective metric | Subjective metric | Disentangled features | Conditioning type | Main architecture | Auxiliary architecture | Application | Goal | Sample rate |
|---|---|---|---|---|---|---|---|---|---|---|---|---|---|---|---|---|---|---|---|
| ACVAE-VC: Non-Parallel Voice Conversion With Auxiliary Classifier Variational Autoencoder | 2019 | C-VAE | Unsupervised | WORLD | n | VAE loss, classification loss | - | - | VCC 2018 | - | - | ABX | Content, spectral envelope, pitch | Parameter | Gated CNN | Class label | Voice conversion | Disentangle content and speaker | 22,05 kHz |
| Voice Conversion With CycleRNN-Based Spectral Mapping and Finely Tuned WaveNet Vocoder | 2019 | Cyclic AE | Unsupervised | WaveNet | n | Reconstruction loss, cycle consistency loss | Acoustic | - | VCC 2018, CMU-ARCTIC | Adam | MCD, LGD | MOS | Content, spectral envelope, pitch | Parameter | RNN | - | Voice conversion | Improve output | 22,05 kHz |
| Non-parallel Voice Conversion using Generative Adversarial Networks | 2018 | AE | Unsupervised | - | y | Feature discriminator loss | - | Freq | VCC 2016 | RMSProp | MCD | ABX | Spectral envelope, mel-cepstrum | General | - | - | Voice conversion | Improve voice quality | - |
| CycleGAN-VC: Non-parallel Voice Conversion Using Cycle-Consistent Adversarial Networks | 2018 | CycleGAN | Unsupervised | Generator | y | Adversarial loss, cycle consistency loss, identity mapping loss | Distribution | - | VCC 2016 | Adam | RMSE, MS | MOS | Content | Parameter | Gated CNN | - | Voice conversion | Improve output | 16 kHz |
| Multi-target Voice Conversion without Parallel Data by Adversarially Learning Disentangled Audio Representations | 2018 | CAE | Unsupervised | Griffin-Lim | y | Reconstruciton loss, classification loss, adversarial loss | Time | - | VCTK | Adam | GV | MOS | Content, speaker | General | CNN + BiLSTM | CNN | Voice conversion | Disentangle content and speaker | 16 kHz |
| RHYTHM-FLEXIBLE VOICE CONVERSION WITHOUT PARALLEL DATA USING CYCLE-GAN OVER PHONEME POSTERIORGRAM SEQUENCES | 2018 | CycleGAN | Unsupervised | Griffin-Lim | y | Adversarial loss, cycle consistency loss, identity mapping loss | Distribution | Distribution | Librispeech, VCTK | - | - | MOS, MOS | Content, rhythm | General | CNN + GRU | - | Style conversion | Improve output | - |
| Frame Selection in SI-DNN Phonetic Space with WaveNet Vocoder for Voice Conversion without Parallel Training Data | 2018 | DNN | Unsupervised | WaveNet, LPC vocoder | n | - | - | - | WSJ CSR CORPUS, SI-284 | - | LSD | AB | Content | General | ASR | - | Voice conversion | Spectral conversion | 16 kHz |
| Voice Conversion with Conditional SampleRNN | 2018 | A-S | Unsupervised | WaveNet | n | Negative Log Likelihood | Time | - | CMU-ARCTIC | Adam | - | - | Content, speaker, pitch | Parameter | ASR | - | Voice conversion | Improve Output | 16 kHz |
| WaveNet Vocoder with Limited Training Data for Voice Conversion | 2018 | CAE | Unsupervised | WaveNet | n | - | - | - | VCC 2018 | Adam | - | MOS | Content, speaker, pitch | General | LSTM | CNN | Voice conversion | Improve output | 16 kHz |
| Investigation of using disentangled and interpretable representations for one-shot cross-lingual voice conversion | 2018 | Seq2seq | Unsupervised | WORLD | n | VAE loss | - | - | TIMIT | Adam | - | CMOS | Content, speaker | General | LSTM | - | Cross-lingual conversion | Improve output | 16 kHz |
| Sequence-to-Sequence Voice Conversion with Similarity Metric Learned Using Generative Adversarial Networks | 2017 | Seq2seq | Unsupervised | - | y | Adversarial loss, similarity loss | Distribution | Acoustic | - | Adam | - | AB | Content, pitch | General | Gated CNN | - | Accent conversion | Improve output | 16 kHz |
| Spectro-Temporal Modelling with Time-Frequency LSTM and Structured Output Layer for Voice Conversion | 2017 | AE | Unsupervised | STRAIGHT | n | Mean squared error (MSE) | Acoustic | - | CMU-ARCTIC | - | MCD | MOS | Timbre, pitch | General | LSTM | - | Voice conversion | Improve Output | 16 kHz |
| Fast Many-to-One Voice Conversion using Autoencoders | 2017 | AE | Unsupervised | - | n | Reonstruction loss | Acoustic | - | - | Adam | LSD | MOS | Content, spectral envelope, pitch | Parameter | MLP | - | Voice conversion | Low-latency | - |



\end{document}